\documentclass[preprint,12pt]{elsarticle}

\usepackage{graphicx, xcolor}
\usepackage[utf8]{inputenc}

\usepackage{amsmath, amsfonts, amssymb, amsthm}
\numberwithin{equation}{section}
\usepackage[subpreambles=false]{standalone}

\usepackage{etoolbox}
\usepackage{url}

\usepackage{breqn}
\usepackage{import}
\usepackage{subcaption}
\usepackage[ruled,vlined,algosection]{algorithm2e}
\usepackage[font=small,labelfont=bf]{caption}
\usepackage[toc,page]{appendix}
\usepackage{placeins}
\usepackage{ifthen}
\usepackage{anyfontsize}
\usepackage{tikz}
\usepackage[inline]{enumitem}

\usepackage{hyperref}
\hypersetup{
    hidelinks,
    colorlinks=true,
    citecolor=green,
    linkcolor=green,
}
\usepackage[noabbrev,capitalize,nameinlink]{cleveref}

\usetikzlibrary{decorations.pathreplacing, shapes.geometric, calc, decorations.pathmorphing, arrows.meta}

\ifpdf
  \DeclareGraphicsExtensions{.eps,.pdf,.png,.jpg}
\else
  \DeclareGraphicsExtensions{.eps}
\fi

\DeclareMathOperator{\sign}{sgn}

\newtheorem{definition}{Definition}[section]
\newtheorem{observation}[definition]{Observation}
\newtheorem{proposition}[definition]{Proposition}

\newtheorem{lemma}[definition]{Lemma}
\newtheorem{theorem}[definition]{Theorem}
\newtheorem{corollary}[definition]{Corollary}

\SetKwInput{kwInit}{Init}
\SetKw{Or}{\hspace{\algoskipindent}\itshape or\hspace{\algoskipindent}}
\SetKw{And}{\hspace{\algoskipindent}\itshape and\hspace{\algoskipindent}}

\graphicspath{{./clipart/}}

\journal{Theoretical Computer Science}

\ifpdf
\hypersetup{
  pdftitle={Optimal Physical Sorting of Mobile Agents},
  pdfauthor={Dmitry Rabinovich, Michael Amir and Alfred M. Bruckstein}
}
\fi

\begin{document}

\begin{frontmatter}
    




\title{Optimal Physical Sorting of Mobile Agents}

\author[Technion]{Dmitry Rabinovich}
\ead{dmitry.ra@cs.technion.ac.il}
\author[Technion]{Michael Amir}
\ead{ammicha3@cs.technion.ac.il}
\author[Technion]{Alfred M. Bruckstein}
\ead{freddy@cs.technion.ac.il}
\affiliation[Technion]{organization={Technion - Israel Institute of Technology},
            addressline={Department of Computer Science}, 
            city={Haifa},
            postcode={3200000}, 
            country={Israel}}

\begin{abstract}
Given a collection of red and blue mobile agents located on two grid rows, we seek to move all the blue agents to the far left side and all the red agents to the far right side, thus \textit{physically sorting} them according to color. The agents all start on the bottom row. They move simultaneously at discrete time steps and must not collide. Our goal is to design a centralized algorithm that controls the agents so as to sort them in the least number of time steps. 

We derive an \textit{exact} lower bound on the amount of time any algorithm requires to sort a given initial configuration of agents. We find an instance optimal algorithm that provably matches this lower bound, attaining the best possible sorting time for any initial configuration. Surprisingly, we find that whenever the leftmost agent is red and the rightmost agent is blue, a straightforward decentralized and local sensing-based algorithm is at most $1$ time step slower than the centralized instance-optimal algorithm.
\end{abstract}



\begin{keyword}
swarm robotics, physical sorting, multi-agent systems, mobile vehicles, smart transportation

\MSC 68T40 \sep 68W40 \sep 68W15 \sep 90B20


\end{keyword}

\end{frontmatter}

\section{Introduction}
Suppose a number of mobile agents are moving on a row. Some of the agents need to travel left, and the other agents need to travel right to arrive at their destination. The agents are not allowed to collide, but have access to another adjacent, initially empty row that they can use to manoeuvre past each other. What is the most efficient way for the agents to achieve their goal and arrive at their desired left-side or right-side destinations?

\begin{figure}
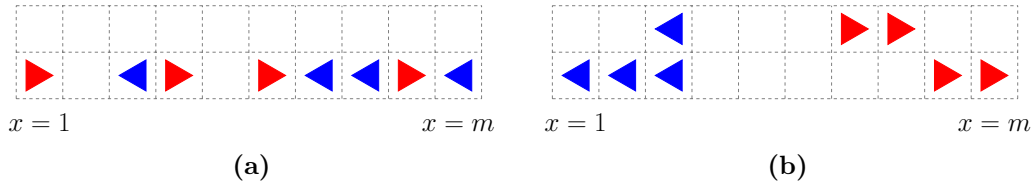

    \centering
        \begin{subfigure}{.48\textwidth}
			\resizebox{\textwidth}{!}{	\begin{tikzpicture}
	    \pgfmathsetmacro{\steplength}{1.5}
	    \pgfmathsetmacro{\size}{10}
	    \begin{scope}[scale=\steplength]
    	    \draw[dashed, black!50] (0, 0) grid (\size, 2);
    	    \begin{scope}[shift={({-0.5 + sqrt(3) / 6}, {1/2-1/3})}]
        	    \foreach \x in {3, 7, 8, 10}
        	    {
        	        \begin{scope}[xshift = \x cm, scale = 2/3]
                        \import{images/}{blocks.3.blue}
                    \end{scope}
                }
            \end{scope}
    	    \begin{scope}[shift={({0.5 - sqrt(3) / 6}, {1/2-1/3})}]
        	    \foreach \x in {0, 3, 5, 8}
        	    {
        	        \begin{scope}[xshift = \x cm, scale=2/3]
                        \import{images/}{blocks.3.red}
                    \end{scope}
                }
            \end{scope}
            \node[anchor=base] at (0 + 0.5, -0.7) {\Huge $x=1$ };
            \node[anchor=base] at (\size - 1 + 0.5, -0.7) {\Huge $x=m$ };
        \end{scope}
	\end{tikzpicture}}
			\caption{}
			\label{sub:initial.config}
		\end{subfigure}
		\hfill
        \begin{subfigure}{.48\textwidth}
			\resizebox{\textwidth}{!}{	\begin{tikzpicture}
	    \pgfmathsetmacro{\steplength}{1.5}
	    \pgfmathsetmacro{\size}{10}
	    \begin{scope}[scale=\steplength]
    	    \draw[dashed, black!50] (0, 0) grid (\size, 2);
    	    \begin{scope}[shift={({-0.5 + sqrt(3) / 6}, {1/2-1/3})}]
        	    \foreach \x/\y in {1/0, 2/0, 3/1, 3/0}
        	    {
        	        \begin{scope}[shift = {(\x, \y)}, scale = 2/3]
                        \import{images/}{blocks.3.blue}
                    \end{scope}
                }
            \end{scope}
    	    \begin{scope}[shift={({0.5 - sqrt(3) / 6}, {1/2-1/3})}]
        	    \foreach \x/\y in {6/1,7/1,8/0,9/0}
        	    {
        	        \begin{scope}[shift = {(\x, \y)}, scale=2/3]
                        \import{images/}{blocks.3.red}
                    \end{scope}
                }
            \end{scope}
            \node[anchor=base] at (0 + 0.5, -0.7) {\Huge $x=1$ };
            \node[anchor=base] at (\size - 1 + 0.5, -0.7) {\Huge $x=m$ };
        \end{scope}
	\end{tikzpicture}}
			\caption{}
			\label{sub:final.config}
		\end{subfigure}
    \caption{(\subref{sub:initial.config}) illustrates an initial configuration of agents. (\subref{sub:final.config}) illustrates a sorted configuration (there are many possible such configurations).}
    \label{fig:typical.configs}
\end{figure}

In this work we study a discrete formalization of this problem. Given a collection of red and blue mobile agents located on parallel grid rows of equal length, we seek a centralized algorithm to move all the blue agents to the far left side and all the red agents to the far right side columns, thus \textit{sorting} the agents  according to color (see \cref{fig:typical.configs}). We assume all agents are initially located on the bottom row. Agents can move simultaneously at discrete time steps $T=0, 1, \ldots$, and must not collide with each other (two agents may never occupy or attempt to move to the same location). Our goal is to design a centralized algorithm that controls the agents so as to sort them in the least number of time steps. 

\begin{figure}[!ht]
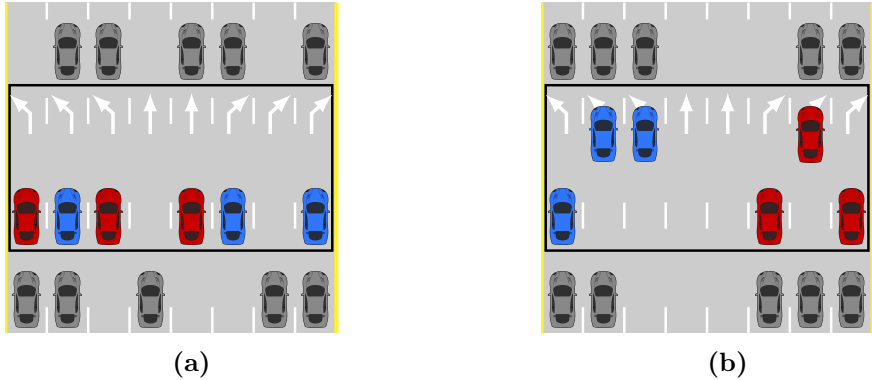

        \begin{subfigure}{.48\textwidth}
            \centering
            \scalebox{1.1}{\import{./images/}{road.frame.cars.only}}
            \caption{}
            \label{sub:unsorted.cars}
        \end{subfigure}
        \hfill
        \begin{subfigure}{.48\textwidth}
            \centering
            \scalebox{1.1}{\import{./images/}{road.frame.cars.sorted}}
            \caption{}
            \label{sub:sorted.cars}
        \end{subfigure}
    \caption{Vehicles driving on a freeway. The blue vehicles want to exit the freeway via an upcoming left exit (not illustrated), and the red vehicles want to exit the freeway via an upcoming right exit. Vehicles need to shift their position to the left or right ahead of time to prepare for exiting the freeway. (\subref{sub:unsorted.cars}) shows an unsorted configuration, and (\subref{sub:sorted.cars}) shows the sorted configuration, after the vehicles have adjusted their positions.}
    \label{fig:car_frame}
\end{figure}

This problem first arose as part of the authors' ongoing research into traffic management algorithms for self-driving vehicles on a freeway. In the context of traffic management, the two rows represent a moving subsection of an upwards-facing freeway. The subsection tracks a set of vehicles all driving at the same forward velocity. We assume that freeway exits might be placed to the left or right of the road. Thus, it is necessary to shift all vehicles that need to exit the freeway leftwards or rightwards ahead of time depending on their desired exit direction, i.e., to ``sort'' them (see \cref{fig:car_frame}). In this work we focus on the special case where the top row is initially empty, which we believe to be independently interesting. 

Similar problems in vehicular control, warehouse management, and combinatorial puzzles have been investigated in the literature and might broadly be  referred to as ``physical sorting'' problems \cite{zhou2016asymmetric, krupke2015parallel, litus2010fall, shang2016approach, shang2014swarm,ratner1990n2,johnson1879notes}. In physical sorting problems, a collection of mobile agents that occupy physical space must bypass each other without colliding in order to arrive at some predefined sorted configuration.

\begin{figure}[!ht]
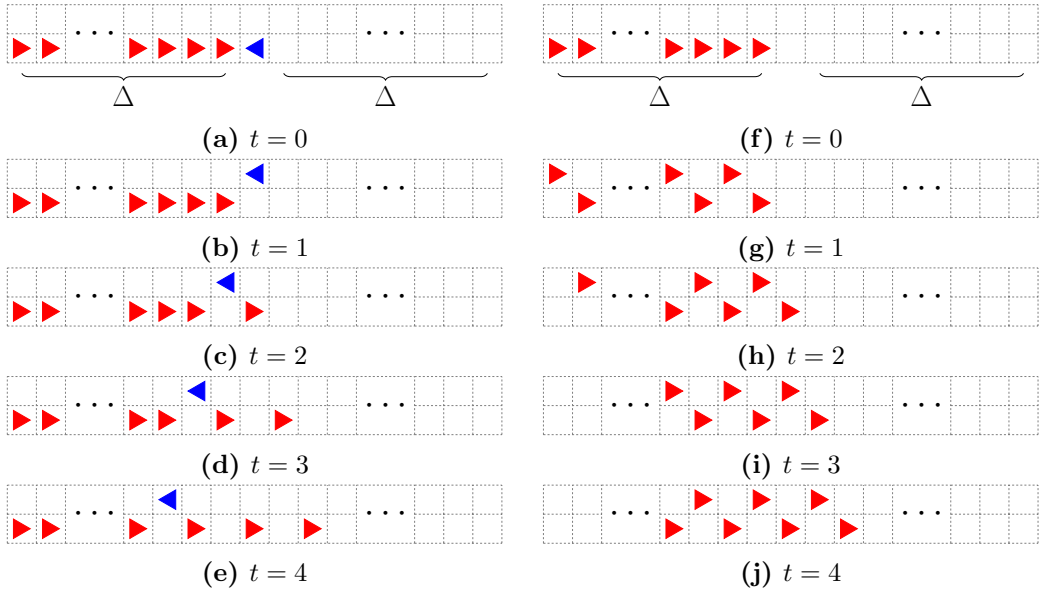

    \centering
    \begin{minipage}{0.48\textwidth}
        \foreach \tick in {0, ..., 4}
        {
        \begin{subfigure}{\textwidth}
			\resizebox{\textwidth}{!}{	\begin{tikzpicture}
	    \pgfmathsetmacro{\steplength}{1.5}
	    \pgfmathsetmacro{\deltaSize}{8}
	    \pgfmathsetmacro{\size}{2 * \deltaSize + 1}
	    \begin{scope}[scale=\steplength]
    	    \draw[dashed, black!50] (0, 0) grid (2, 2);
    	    \draw[dashed, black!50] (4, 0) grid (\size - 5, 2);
    	    \draw[dashed, black!50] (\size - 3, 0) grid (\size, 2);
    	    \foreach \x in {2}
    	    {
        	    \foreach \y in {0, 2}
        	    {
            	    \draw[dashed, black!50] (\x, \y) -- (\x + 2, \y);
            	    \draw[dashed, black!50] (\size - 7 + \x, \y) -- (\size - 5 + \x, \y);
        	    }
    	    }
    	    \begin{scope}[shift={({-0.5 + sqrt(3) / 6}, {1/2-1/3})}]
        	    \foreach \x in {\deltaSize}
        	    {
        	        \pgfmathsetmacro{\xposition}{\x + 1 - (\tick - 1) * (\tick > 1)}
        	        \begin{scope}[shift = {(\xposition, \tick > 0)}, scale = 2/3]
                        \import{images/}{blocks.3.blue}
                    \end{scope}
                }
            \end{scope}
    	    \begin{scope}[shift={({0.5 - sqrt(3) / 6}, {1/2-1/3})}]
        	    \foreach \x in {0, ..., 2}
        	    {
        	        \pgfmathsetmacro{\xposition}{\x + 2 * (\x == 2)}
        	        \begin{scope}[xshift = \xposition cm, scale=2/3]
                        \import{images/}{blocks.3.red}
                    \end{scope}
        	        \pgfmathsetmacro{\xposition}{\deltaSize - \x - 1 + (\tick - \x - 1) * (\x + 1 < \tick)}
        	        \begin{scope}[xshift = \xposition cm, scale=2/3]
                        \import{images/}{blocks.3.red}
                    \end{scope}
                }
            \end{scope}
            \node at (3, 1) {\fontsize{54}{12}\selectfont $\cdots$};
            \node at (2 * \deltaSize - 3, 1) {\fontsize{54}{12}\selectfont $\cdots$};
            \begin{scope}[shift={(0.5, 0.5)}]
                \ifthenelse{\tick=0}
                {
                    \draw [decorate,decoration={brace,mirror, amplitude=10pt},xshift=0pt,yshift=4pt]
                        (0, -1) -- (\deltaSize - 1, -1) node [black,midway,yshift=-1.2cm] 
                        {\fontsize{42}{12}\selectfont $\Delta$};
                    \draw [decorate,decoration={brace,mirror, amplitude=10pt},xshift=0pt,yshift=4pt]
                        (\deltaSize + 1, -1) -- (2 * \deltaSize, -1) node [black,midway,yshift=-1.2cm] 
                        {\fontsize{42}{12}\selectfont $\Delta$};
                }
                {}
            \end{scope}
        \end{scope}
	\end{tikzpicture}}
			\caption{$t=\tick$}
			\label{sub:Example.2Delta.\tick}
		\end{subfigure}
		}
	\end{minipage}
	\hfill
    \begin{minipage}{0.48\textwidth}
        \foreach \tick in {0, ..., 4}
        {
        \begin{subfigure}{\textwidth}
			\resizebox{\textwidth}{!}{	\begin{tikzpicture}
	    \pgfmathsetmacro{\steplength}{1.5}
	    \pgfmathsetmacro{\deltaSize}{8}
	    \pgfmathsetmacro{\size}{2 * \deltaSize + 1}
	    \begin{scope}[scale=\steplength]
    	    \draw[dashed, black!50] (0, 0) grid (2, 2);
    	    \draw[dashed, black!50] (4, 0) grid (\size - 5, 2);
    	    \draw[dashed, black!50] (\size - 3, 0) grid (\size, 2);
    	    \foreach \x in {2}
    	    {
        	    \foreach \y in {0, 2}
        	    {
            	    \draw[dashed, black!50] (\x, \y) -- (\x + 2, \y);
            	    \draw[dashed, black!50] (\size - 7 + \x, \y) -- (\size - 5 + \x, \y);
        	    }
    	    }
    	    \begin{scope}[shift={({0.5 - sqrt(3) / 6}, {1/2-1/3})}]
        	    \foreach \x in {0, ..., 5}
        	    {
        	        \pgfmathsetmacro{\xboost}{2 * ((\x > 2 - \tick) + (\tick == 0) * (\x == 2)) + (\tick - 1) * (\tick > 1)}
        	        \pgfmathsetmacro{\yboost}{(\tick > 0) * (1 - mod(\x, 2)) }
        	        \begin{scope}[shift = {(\x + \xboost, \yboost)}, scale=2/3]
                        \import{images/}{blocks.3.red}
                    \end{scope}
                }
                \foreach \x in {4, ..., 11}
                {
                    \ifthenelse{\x<\tick \OR \x=\tick}
                    {
            	        \pgfmathsetmacro{\yboost}{mod(\tick + \x, 2)}
            	        \begin{scope}[shift = {(\x, \yboost)}, scale=2/3]
                            \import{images/}{blocks.3.red}
                        \end{scope}
                    }{}
                }
            \end{scope}
            
            \node at (3, 1) {\fontsize{54}{12}\selectfont $\cdots$};
            \node at (2 * \deltaSize - 3, 1) {\fontsize{54}{12}\selectfont $\cdots$};
                \begin{scope}[shift={(0.5, 0.5)}]
                    \ifthenelse{\tick=0}
                    {
                        \draw [decorate,decoration={brace,mirror, amplitude=10pt},xshift=0pt,yshift=4pt]
                            (0, -1) -- (\deltaSize - 1, -1) node [black,midway,yshift=-1.2cm] 
                            {\fontsize{42}{12}\selectfont $\Delta$};
                        \draw [decorate,decoration={brace,mirror, amplitude=10pt},xshift=0pt,yshift=4pt]
                            (\deltaSize + 1, -1) -- (2 * \deltaSize, -1) node [black,midway,yshift=-1.2cm] 
                            {\fontsize{42}{12}\selectfont $\Delta$};
                    }
                    {}
                \end{scope}
        \end{scope}
	\end{tikzpicture}}
			\caption{$t=\tick$}
			\label{sub:Example.Delta+1.\tick}
		\end{subfigure}
		}
	\end{minipage}
    \caption{(\subref{sub:Example.2Delta.0})-(\subref{sub:Example.2Delta.4}) illustrate the first five time steps of an optimal strategy for sorting a normal initial configuration with a single blue agent placed in front of a red agent group. No sorting is possible in faster than $2\Delta$ ticks. Subfigures 
(\subref{sub:Example.Delta+1.0})-(\subref{sub:Example.Delta+1.4}) illustrate non-normal initial configuration that can be sorted in $\Delta + 1$ ticks via splitting the agents between the rows in an alternating fashion (\cref{algo:general.movement}). The first five time steps of this splitting strategy are shown.
Both initial configurations share a very long sequence of red agents. The only difference is a lone blue agent. However, the optimal sorting times are strikingly different.}
    \label{fig:2Delta images}
\end{figure}

A trivial algorithm that accomplishes our agents' sorting task is the following: move all the agents of one color (say, red) to the top row; then let the red  agents  move right and the blue agents  move left in their respective row (\cref{fig:2Delta images}, a-e). This algorithm does not require complex computation nor even centralized decision-making--it is a straightforward, decentralized, local strategy that can be executed by simple autonomous agents without requiring any global knowledge on the agent configuration.

How does the above simple algorithm fare compared to an optimal centralized sorting algorithm? A priori, since the number of possible strategies for the agents is enormous, one would expect far better sorting strategies are available. However, in this work, we shall prove the  surprising result that this ``trivial'' distributed strategy is at most one time step short of optimal for a very large class of initial agent configurations called \textit{normal configurations} (configurations where the leftmost agent wants to go right and the rightmost agent wants to go left), and is in fact optimal over such configurations assuming we choose the correct color to move to the spare row. 

In the general case where we also consider non-normal configurations, we show that  the optimal makespan is determined by the maximal  normal subconfiguration, and find a provably optimal sorting algorithm for the agents  (\cref{NonNormalConfigurationSection}). Furthermore, we derive an \textbf{exact} lower bound for the amount of time it takes to sort the agents given any starting configuration (\cref{theorem:makespanresult}). Our proposed optimal algorithm matches this lower bound, thus it is instance optimal in the sense of \cite{afshani2017instance}, attaining the best possible sorting time for any initial configuration.

The algorithm (\cref{algo:general.movement}) can be understood as blending two strategies: inside the maximal normal subconfiguration, we split the agents into rows based on their desired direction of motion, according to the aforementioned ``trivial'' strategy. Outside this subconfiguration, agents split between both rows to move faster regardless of their color (\cref{figure:subnormalalgoexample}). The idea behind this algorithm is simple to understand, but the implementation requires several delicate caveats which are discussed throughout this work.

Physical sorting can sometimes be unintuitive: the addition of a single agent can completely change the optimal strategy and double the time to completion of the sorting (\cref{fig:2Delta images}). Furthermore, when attempting to derive lower bounds on the makespan, the infinite set of strategies available to the agents makes it difficult to keep track of each agent, which greatly complicates the mathematical proofs. We overcome these complexities by identifying a small set of \textit{critical agents} (see \cref{definition:critical-agents}) and showing, roughly, that only the movements of the critical agents at a given time step can affect any algorithm's makespan. 

\begin{figure}
    \centering
    \foreach \tick in {0, ..., 4}
    {
    \begin{subfigure}{0.7\textwidth}
		\resizebox{\textwidth}{!}{	\begin{tikzpicture}
	    \pgfmathsetmacro{\steplength}{1.5}
	    \pgfmathsetmacro{\size}{14}
	    \begin{scope}[scale=\steplength]
    	    \draw[dashed, black!50] (0, 0) grid (\size, 1);
    	    \begin{scope}[shift={({0.5 - sqrt(3) / 6}, {1/2-1/3})}]
        	    \foreach \x[count=\i] in {0, 1, 3, 4, 7, 11, 12}
        	    {
        	        \global\let\agentnum=\i
        	        \expandafter\xdef\csname x\i\endcsname{\x}
                }
                \ifthenelse{\tick > 0}{
                    \foreach \ttick in {1, ..., \tick}
        	        {
        	            \foreach \num in {2, ..., \agentnum}
        	            {
        	                \expandafter\let\expandafter\tempa\csname x\num\endcsname
        	                \pgfmathtruncatemacro{\prev}{\num - 1}
        	                \expandafter\let\expandafter\tempb\csname x\prev\endcsname
        	                \pgfmathtruncatemacro{\compareto}{\tempa-1}
        	                \ifthenelse{\tempb<\compareto}
        	                {  
            	                \pgfmathtruncatemacro{\gotox}{\tempb + 1}
                    	        \global\expandafter\let\csname x\prev\endcsname=\gotox
        	                }
        	                {}
        	            }
                        \expandafter\let\expandafter\tempb\csname x\agentnum\endcsname
    	                \pgfmathtruncatemacro{\compareto}{\size-1}
        	            \ifthenelse{\tempb<\compareto}
        	            {
            	                \pgfmathtruncatemacro{\gotox}{\tempb + 1}
                    	        \global\expandafter\let\csname x\agentnum\endcsname=\gotox
        	            }{}
        	        }
                }{}
        	    \foreach \num in {1, ..., \agentnum}
        	    {
    	            \expandafter\let\expandafter\tempa\csname x\num\endcsname
        	        \begin{scope}[shift = {(\tempa, 0)}, scale=2/3]
                        \import{images/}{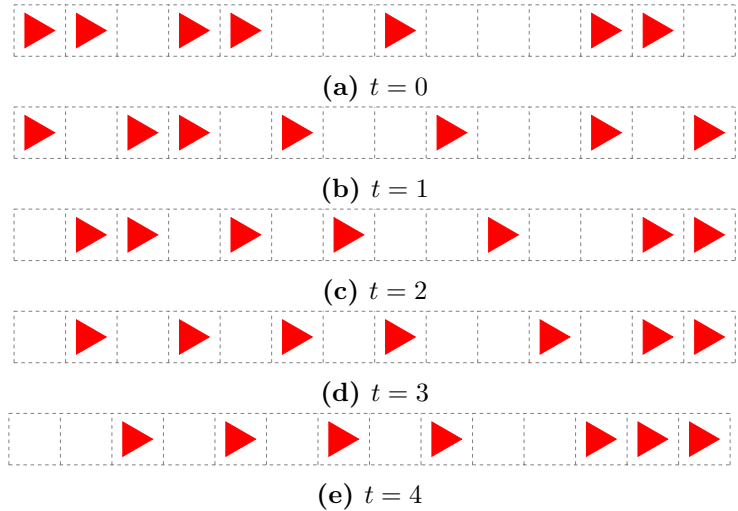}
                    \end{scope}
                }
           \end{scope}
        \end{scope}
	\end{tikzpicture}}
		\caption{$t=\tick$}
		\label{sub:Example.OneRow.\tick}
	\end{subfigure}
	}
    \caption{Subfigures (\subref{sub:Example.OneRow.0})-(\subref{sub:Example.OneRow.4}) illustrate a single row  model with agents moving in one direction. The model could be viewed as a TASEP with $p_{\text{right}} = 1$ and $p_{\text{left}}=0$. In the provided example, all agents reach the far-right side on the $9$th time step, which is equal to what we define as the $f_{max}$ value of the configuration (see \cref{def:f} and \cref{corollary:singlerowtime})} 
    \label{fig:One row TASEP - like}
\end{figure}

We believe that the expression for the lower bound obtained in \cref{theorem:makespanresult} is  independently interesting. As an example application, \cref{corollary:singlerowtime}, which is a small special case of the expression, computes the amount of time it takes for a ``traffic jam'' of agents on a single row to get to the right (left) side of the row, assuming each agent moves right (left) at every time step where there is no agent in front of them (see \cref{fig:One row TASEP - like}). This corollary relates to the study of the \textit{totally asymmetric simple exclusion process} in statistical mechanics, which has been used to study transport phenomena such as traffic flow and biological transport \cite{chou2011biologicaltasep,kriecherbauer2010pedestrian,chowdhury2000statistical}. The lower bound of \cref{theorem:makespanresult} can be applied to compute the time it takes an arbitrary TASEP particle configuration with synchronized waking times to get from one end of the row to the other, assuming $p_{right} = 1$ (probability of moving right) and $p_{left} = 0$. Although this corollary is fairly straightforward, we could not find a similar result in the TASEP literature. 

\section{Related Work}
Various ``physical sorting'' problems appear in the literature \cite{zhou2016asymmetric, krupke2015parallel, litus2010fall, shang2016approach, shang2014swarm} with applications to warehouse logistics, where loads must be efficiently moved to different ends of a warehouse (in particular, our assumptions  resemble those of puzzle based storage systems \cite{gue2007puzzlebasedstorage}); servicing, where mobile robots self-sort according to some priority ordering; assembling, where tasks should be carried in a predefined order which respects physical space; and transportation. In such problems, mobile agents (robots, vehicles, etc.) that take up physical space are required to attain some kind of sorted configuration while avoiding collisions and maneuvering around each other, making their decisions in either a distributed \cite{zhou2016asymmetric} or centralized \cite{petig2018changing} fashion. For example, similar to this work, in \cite{petig2018changing}, the authors consider a task of sorting mobile vehicles in two rows, developing approximation algorithms and proving computational hardness results. Their goal, however, is to \textit{split} the vehicles between the rows according to their color as fast as possible, as opposed to our goal of bringing robots to the left-hand or right-hand side depending on their color. 

The famous ``15 puzzle", where numbered tiles must be slid across a grid until they are ordered, can also be seen as a physical sorting task for which algorithms and computational hardness results are available  \cite{johnson1879notes,ratner1990n2}. 

Lane changing has been intensively studied in the past in many traffic models \cite{hatipoglu1995optimal,chee1994vehicle,naranjo2008lane,atagoziyev2016lane}, with recent results in both centralized and distributed control protocols \cite{visintainer2016towards, ekenstedt2019membership,schubert2010situation}.

As previously mentioned, a side application of our results (\cref{corollary:singlerowtime})  relates to \textit{totally simple exclusion processes} (TASEPs) in statistical mechanics  \cite{chou2011biologicaltasep,kriecherbauer2010pedestrian,chowdhury2000statistical}. TASEPs have been used, in particular, as an idealized model that captures many of the phenomena of real-world vehicular traffic. The general model we present in this work is, however, not directly related to any TASEP model, since we study deterministic mobile agents that act according to an intelligent, centralized algorithm, rather than a predetermined stochastic process.

Broadly speaking, our mathematical modelling of the problem (\cref{section:model}) also relates to various discrete grid-like settings in multi-robot and multi-agent systems, wherein a large number of robots whose spatial locations are represented as coordinates on a grid-like region all move synchronously according to a local or centralized algorithm while avoiding collisions. The goal of agents in these settings can be, for example, fast deployment, gathering, or formation on an a priori unknown grid environment \cite{arxivminimizingtravel,uniformdispersion,barraswarm1,altshuler2018introduction,siamcomputing_dieudonne2015meetagentsasynchronously,siamcomputing_fujinaga2015patternmobilerobot,hideg2020asynchronous}.






\section{Model}
\label{section:model}

We are given $n$ mobile agents on a $2 \times m$ grid environment, such that $n_1$ agents are red and $n_2$ agents are blue ($n_1 + n_2 = n$). Every agent begins in an $(x,y)$  coordinate of the form $(\cdot, 1)$ (see \cref{sub:initial.config}), starting at $x=1$. A column of the environment is called \textit{red-occupied} if it contains at least one red agent and no blue agents, \textit{blue-occupied} if it contains at least one blue agent and no red agents, \textit{mixed} if it contains both colors of agents, and \textit{empty} otherwise. The goal of the agents is to move to a configuration such that the red-occupied columns are the rightmost columns, and the blue-occupied columns are the leftmost columns. Formally, the following conditions must be fulfilled:

\begin{enumerate}
    \item There are no mixed columns.
    \item Every blue-occupied column is to the left of every red-occupied column and every empty column.
    \item Every red-occupied column is to the right of every blue-occupied column and every empty column.
\end{enumerate}

When the agents achieve such a configuration we say the system is \textit{sorted} (see \cref{sub:final.config}), as it separates all agents in the configuration to the far left or far right columns of the environment based on color. It might be desirable to require, in addition to the above three conditions, that all agents end up on the bottom row (just as they began): in all algorithms we present here, this can be achieved at the cost of at most one additional time step that moves all the robots downwards, and analogous optimality results hold. 

We define a model of agent motion based on common assumptions in synchronous and semi-synchronous models of traffic flow \cite{chou2011biologicaltasep}, biological systems \cite{amir2022locust,amir2020discretelocusts}, and the theory of mobile multi-robot systems \cite{arxivminimizingtravel,uniformdispersion,barraswarm1,altshuler2018introduction,siamcomputing_dieudonne2015meetagentsasynchronously,siamcomputing_fujinaga2015patternmobilerobot}. 

Time is discretized into steps $t = 0, 1, \ldots$. At every time step, agents may move to adjacent empty locations (up, down, left, or right, non-diagonally). Agents cannot move to a location that already contains an agent, nor can two agents move to the same empty location at the same time. 

Note that, according to the above assumptions, a robot can only move into a location that is unoccupied, but two or more adjacent agents followed by an empty location cannot both move right (or left) in the same time step, meaning that such situations result in a traffic jam or ``shock'': the agents at the leftmost (rightmost) end of the line must wait several time steps for empty space to be created between them and the rest of the agents (see the first column of \cref{fig:2Delta images}). This is a common assumption in the literature on traffic flow, reflecting the fact that adjacent agents (even those receiving commands transmitted by a central algorithm) do not have perfectly coordinated clocks nor perfect motion detection systems and thus cannot begin moving at the exact same time and at the exact same speed without risking collisions.

The \textit{beginning} of a time step refers to the configuration before any agent movements, and the \textit{end} of the time step refers to the configuration after agent movements (so the configuration at the beginning of time $t+1$ is the same as at the end of time $t$). Unless explicitly stated otherwise, anywhere in this work, when we refer to the agents' configuration ``at time $t$", we mean the configuration at the beginning of that time step.

The agents' actions are controlled by a sorting algorithm. The number of time steps it takes an algorithm to move the agents from the initial configuration to a sorted configuration is called the \textit{makespan} of that algorithm with respect to the initial configuration (i.e., the makespan is the first $t$ such that the configuration is sorted at the beginning of time step $t$).

\begin{definition}\label{def:normal-configuration}
A \textbf{normal} initial configuration is an initial configuration of agents such that the rightmost agent is blue and the leftmost agent is red.
\end{definition}

Unless stated otherwise, we assume in this work that the initial agent configuration is normal. We will specifically handle non-normal initial configurations in \cref{NonNormalConfigurationSection}.  Certain strategies, such as an alternating split between the rows that enables each agent to move horizontally faster, are effective in non-normal configurations but ineffective in normal configurations, thus the optimal strategy can be different. An example is shown in \cref{fig:2Delta images}, which compares a normal configuration to a non-normal configuration and shows an optimal strategy for each case. In \cref{fig:2Delta images}, the addition of a single blue agent completely alters the optimal strategy, and doubles the optimal makespan.  

\section{A lower bound on makespan}

In this section, we will prove a lower bound on the makespan of any (centralized or distributed) sorting algorithm for normal configurations. In the next section, we will show this lower bound is tight for normal configurations: there is a centralized algorithm that matches it exactly. 

The lower bound is defined as a function of the initial configuration of agents and empty columns. For every agent $A$, we denote $A$'s location at time $t$ as $(A_x(t), A_y(t))$. Let $\mathcal{A} = \mathcal{R} \cup \mathcal{B}$ be the set of all agents, where $\mathcal{R}$ is the set of red agents and $\mathcal{B}$ is the set of blue agents. 

\begin{definition}
\label{def:frontback}
For any agent $A$, we define $front(A)$ and $back(A)$.
%
%
\begin{enumerate}
    \item If $A \in \mathcal{R}$, then $front(A)$ is the set of all \textit{not}-red-occupied columns (i.e., empty or blue-occupied) to the right right of $A$ (i.e., with strictly greater $x$ coordinate) and $back(A)$ is the set of all red-occupied columns to the left of $A$  (i.e., with strictly smaller $x$ coordinate) at time $0$. 
    \item If $A \in \mathcal{B}$, then $front(A)$ is the set of all \textit{not}-blue-occupied columns to the left of $A$ and $back(A)$ is the set of all blue-occupied columns to the right of $A$ at time $0$.
\end{enumerate}
\end{definition}

\begin{figure}[!ht]
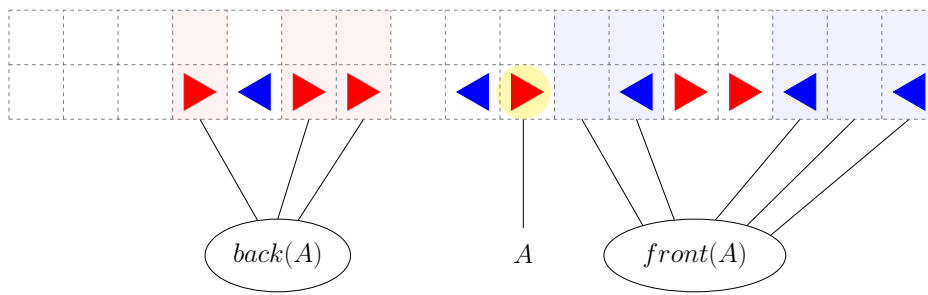

    \centering
		\resizebox{.9\textwidth}{!}{	\begin{tikzpicture}
	    \pgfmathsetmacro{\steplength}{1.5}
	    \pgfmathsetmacro{\toscale}{2/3}
	    \pgfmathsetmacro{\size}{17}
	    \begin{scope}[scale=\steplength]
	        
	        \begin{scope}[shift={({0.5 - sqrt(3) / 24}, 0.5)}]
	            \begin{scope}[xshift=9 cm]
	                \draw[draw = none, fill = yellow!40] (0,0) coordinate (ACenter) circle (0.5cm);
	                \draw ($(ACenter)+(0, -0.5)$) -- ++(0, -2) coordinate (ALabel);
	                \node[below] at ($(ALabel)+(0, -0.2)$) {\LARGE $A$};
	            \end{scope}
	            
                \node[draw,ellipse,minimum height=2cm,minimum width=4cm] (a) at (4.5,-3) {};
	            \node at (4.5, -3) {\LARGE $back(A)$};
	        \end{scope}
            \foreach \back[count=\i] in {3, 5, 6}
            {
                \draw[fill=red!5, draw=none] (\back, 0) coordinate (b\back) rectangle ++(1, 2);
                \draw ($(b\back)+(0.5,0)$) -- (a.{150-30*\i});
            }
	        \begin{scope}[shift={({0.5 + sqrt(3) / 24}, 0.5)}]
                \node[draw,ellipse,minimum height=2cm,minimum width=5cm] (c) at (12,-3) {};
	            \node at (12, -3) {\LARGE $front(A)$};
	        \end{scope}
            \foreach \front/\where[count=\i] in {11/120, 14/60, 16/15}
            {
                \draw[fill=blue!5, draw = none] (\front, 0) coordinate (d\i) rectangle ++(1, 2);
                \draw ($(d\i)+(0.5,0)$) -- (c.\where);
            }
            \foreach \front/\where[count=\i] in {10/150, 15/30}
            {
                \draw[fill=blue!5, draw = none] (\front, 0) coordinate (d\i) rectangle ++(1, 2);
                \draw ($(d\i)+(0.5,0)$) -- (c.\where);
            }

    	    \draw[dashed, black!50] (0, 0) grid (\size, 2);
    	    \begin{scope}[shift={({-0.5 + sqrt(3) / 6}, {1/2-1/3})}]
                \foreach \x[count=\i] in {4, 8, 11, 14, 16}
                {
        	        \begin{scope}[shift = {(\x + 1, 0)}, scale = \toscale]
                        \import{images/}{blocks.3.blue}
                    \end{scope}
                }
            \end{scope}
    	    \begin{scope}[shift={({0.5 - sqrt(3) / 6}, {1/2-1/3})}]
                \foreach \x[count=\i] in {3, 5,6, 9, 12, 13}
                {
        	        \begin{scope}[shift = {(\x, 0)}, scale = \toscale]
                        \import{images/}{blocks.3.red}
                    \end{scope}
                }
            \end{scope}
            
        \end{scope}
	\end{tikzpicture}}
    \caption{The $front$ and the $back$ of the red agent $A$ (highlighted in yellow). We see that $f(A) = 8$  (\cref{def:f}).} 
    \label{fig:example: front and back}
\end{figure}

\cref{def:frontback} is illustrated in \cref{fig:example: front and back}. 

For any agent $A$ we define the value $f(A)$ which is dependent on the initial configuration. We will show that every value $f(A)$ in an initial configuration is a lower bound on the makespan of \textit{any} sorting algorithm executed over that initial configuration. Thus the maximum $f_{max}$ of all these values is a lower bound as well. 

\begin{definition}\label{def:f}
For every agent $A$, we define $f(A)$ as follows (see \cref{fig:example: front and back}):
$$f(A) = |front(A)| + |back(A)|,$$ 

We further define $f_{max} = \max_{A \in \mathcal{A}}f(A)$. 
\end{definition}

\begin{definition} \label{definition:critical-agents}
An agent $A$ for which $f(A) = f_{max}$ is called a \textbf{critical agent}.
\end{definition}

We will show that in many initial configurations, $f_{max}$ is the exact lower bound for the makespan of any sorting algorithm. However, in some situations, such as when the agents are too densely packed together and must waste time without being able to move, it is possible that the algorithm requires one extra time step to sort the agents. To derive an exact lower bound, we will have to take into account this ``plus 1,'' whose preconditions are captured by the following definition:

\begin{definition}
\label{Vmaxdef}
We define $\mathcal{V}$ to equal $1$ if:

\begin{enumerate}
    \item There is both a red and a blue critical agent, or
    \item There is a critical agent $A$ such that, in the initial configuration, another agent is located immediately in front of it (i.e., at $(A_x(0)+1,1)$ if $A$ is red and $(A_x(0)-1,1)$ if $A$ is blue).
\end{enumerate}

Otherwise, $\mathcal{V}=0$.
\end{definition}

The goal of this section and the next one is to prove the following result:

\begin{theorem}
\label{theorem:makespanresult}
The makespan of any physical sorting algorithm for a given normal initial configuration is at least $f_{max} + \mathcal{V}$. This lower bound is precise; there is an algorithm that achieves sorting in exactly $f_{max}+\mathcal{V}$ time steps.
\end{theorem}

For the sake of the proof, it is helpful to track the relative ordering of red and blue agents based on their $x$ coordinate. To this end, we assign each red agent an integer called a  \textit{label}. At time $t=0$, the labels depend on the relative $x$-coordinate position of the agents, such that the \textit{leftmost} red agent has the label $n_1$, the second-leftmost red agent has the label $n_1 - 1$, and so on, with the rightmost red agent having label $1$. Two agents $R$ and $R'$ \textit{exchange} their labels at time $t$ if $R_x(t-1) \leq R_x'(t-1)$ but $R_x(t) > R_x'(t)$ (in other words, they pass each other horizontally). When this happens, $R'$ receives the label $R$ had at time $t-1$, and vice-versa.

We similarly define labels for blue agents, but in the \textit{opposite} order of $x$ coordinates, such that the \textit{rightmost} blue agent receives label $n_2$, and the \textit{leftmost} blue agent receives label $1$. Label exchanging is defined as before.

Note that labels can only be exchanged by agents of the same color. When a pair of red and blue agents pass each other, labels are unchanged.

We define $ord(A, t)$ to equal the agent $A$'s label at time $t$.

Let $\mathcal{P}^B_i(t)$ denote the blue agent $A$ with $ord(A,t) = i$ at time $t$. When $t$ is implicitly obvious, we will simply write $\mathcal{P}^{B}_{i}$. In the text, we will intuitively think of  $\mathcal{P}^{B}_i$ \textbf{as having a ``persistent''  identity}, like an agent, and track its position. We will likewise define $P^R_{i}(t)$ for red agents.

\begin{observation}\label{observation:TagSlowMovement}
    Let $ord(A, t) = ord(A', t + 1)$, where $A$ and $A'$ are agents of the same color. Then $|A_x(t) - A_x'(t+1)| \leq 1$, i.e. a label can not travel faster than one step horizontally in a single time tick.
\end{observation}

\cref{observation:TagSlowMovement} follows from the fact that agents can only ``exchange" labels by bypassing each other. 

We will now prove several claims about blue agents which will establish lower bounds. Symmetrically, all such claims will hold for red agents, which will allow us to wrap up the proof at the end of the section.

We will henceforth assume the agents move according to some makespan-optimal algorithm $\textit{ALG}$. Per \cref{theorem:makespanresult}, our end-goal is to show that $\textit{ALG}$'s makespan is at least $f_{max} + \mathcal{V}$.

\begin{figure}[!ht]
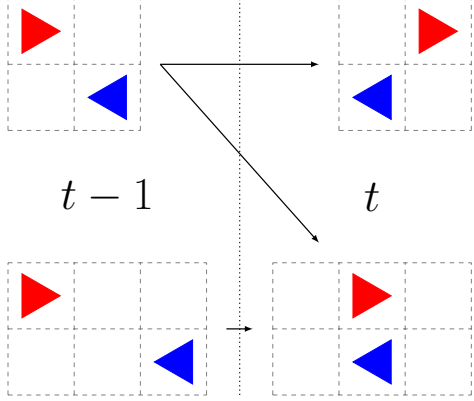

    \centering
		\resizebox{.45\textwidth}{!}{	\begin{tikzpicture}
	    \pgfmathsetmacro{\steplength}{1.5}
	    \begin{scope}[scale=\steplength]
	        \begin{scope}[dashed, black!50]
    	        \draw (0, 0) grid +(3, 2);
    	        \draw (0, 4) grid +(2, 2);
    	        \draw (5, 4) grid +(2, 2);
    	        \draw (4, 0) grid +(3, 2);
	        \end{scope}
	        \begin{scope}[shift={({-0.5 + sqrt(3) / 6}, {1/2-1/3})}]
        	    \foreach \x/\y in {3/0, 2/4, 6/4, 6/0}
        	    {
        	        \begin{scope}[shift = {(\x, \y )}, scale = 2/3]
                        \import{images/}{blocks.3.blue}
                    \end{scope}
                }
            \end{scope}
    	    \begin{scope}[shift={({0.5 - sqrt(3) / 6}, {1/2-1/3})}]
        	    \foreach \x/\y in {0/1, 0/5, 6/5, 5/1}
        	    {
        	        \begin{scope}[shift = {(\x, \y)}, scale=2/3]
                        \import{images/}{blocks.3.red}
                    \end{scope}
                }
            \end{scope}
            \draw[thick, -latex] (3 +0.3, 1) -- (4-0.3, 1);
            \draw[thick, -latex] (2 +0.3, 5) -- (5-0.3, 2+0.3);
            \draw[thick, -latex] (2 +0.3, 5) -- (5-0.3, 5);
            
            \draw[thick, dotted, black!90] (3.5, 0) -- +(0,6);
            \node at (1.5, 3) {\huge $t-1$};
            \node at (5.5, 3) {\huge $t$};
        \end{scope}
	\end{tikzpicture}}
    \caption{The possible outcomes of a ``meeting'' between blue and red agents. On the left are two possible states of the agents at time $t-1$ before meeting, and on the right are the possible states they can transition into after meeting.}
    \label{fig:example.configs}
\end{figure}

\begin{definition}
\label{definition:meeting}
A red agent $R$ and a blue agent $B$ are said to \textbf{meet} at time $t$ if $$\sign{(R_{x}(t-1) - B_{x}(t-1))} \in \{1, -1\}$$ and $$\sign{(R_{x}(t-1) - B_{x}(t-1))} \neq \sign{(R_{x}(t) - B_{x}(t))},$$ where $\sign{(\cdot)}$ denotes the sign ($-1$, $0$ or $1$) of an integer. (See \cref{fig:example.configs}.)

Two labels $P^B$ and $P^R$ are said to meet at time $t$ if their respective agents meet at time $t$.
\end{definition}

\begin{definition}
\label{definition:agentVvalues}
Define $\mathcal{V}_i^{\textit{B}} = 0$ if during the execution of \textit{ALG}, $\mathcal{P}^{B}_i$ decreases its $x$ coordinate at \textit{every} time step before it arrives at column $i$ for the first time \textbf{and} before it meets $\mathcal{P}^{R}_{n_1}$ . Otherwise let $\mathcal{V}_i^{\textit{B}} = 1$. 

Define $\mathcal{V}_i^{\textit{R}} = 0$ if during the execution of \textit{ALG}, $\mathcal{P}^{R}_i$ increases its $x$ coordinate at \textit{every} time step before it arrives at column $n-i+1$ for the first time \textbf{and} before it meets $\mathcal{P}^{B}_{n_2}$. Otherwise let $\mathcal{V}_i^{\textit{R}} = 1$. 
\end{definition}

Informally speaking, the values $\mathcal{V}_i^{\textit{B}}$ and $\mathcal{V}_i^{\textit{R}}$ are a measure of whether a given label ever performs a movement that will propagate backwards and slow down all the labels behind it. We will see how this occurs in \cref{lemma:lower.bound}, but the intuitive idea is this: if there is any time step where, say, the label $\mathcal{P}_i^B$ doesn't move on the $x$ axis towards its final destination (which must be the column $i$ or to the left of it), then every label $P_j^{B}$ with $j > i$ might get obstructed by this movement due to a back-propagating slowdown. For technical reasons apparent in the proof of \cref{lemma:lower.bound}, we want to only factor such movements into our bound if they occur \textit{before}  $\mathcal{P}_i^B$ meets the leftmost red label $\mathcal{P}^{R}_{n_1}$. 

The general plan is to eventually tie \cref{definition:agentVvalues} to the value $\mathcal{V}$ in \cref{Vmaxdef}. 

\begin{figure}
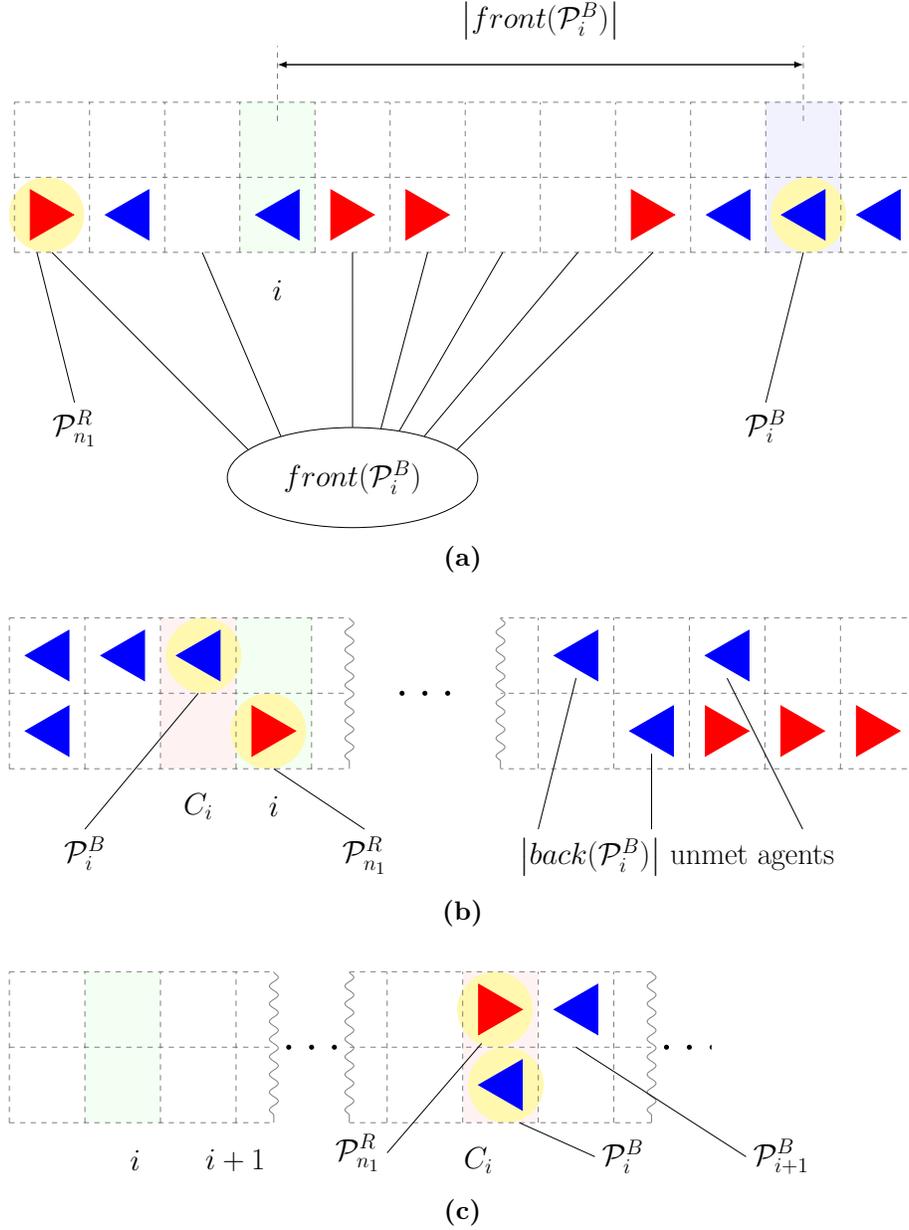

    \centering
    \begin{minipage}{0.88\textwidth}
        \begin{subfigure}{\textwidth}
			\resizebox{\textwidth}{!}{    \begin{tikzpicture}
	    \pgfmathsetmacro{\steplength}{1.5}
	    \pgfmathsetmacro{\toscale}{1 / \steplength}
	    
	    \pgfmathsetmacro{\size}{12}
	    \pgfmathtruncatemacro{\ordinali}{4-1}
	    \pgfmathtruncatemacro{\initiali}{10}
	    
	    \begin{scope}[scale=\steplength]
	        \draw[draw=none, fill = green!5] (\ordinali, 0) rectangle +(1, 2);
	        \draw[draw=none, fill = blue!5] (\initiali, 0) rectangle +(1, 2);

	        \begin{scope}[shift={({0.5 + sqrt(3) / 24}, 0.5)}]
	            \draw[draw = none, fill = yellow!40] (\initiali,0) circle (0.5cm);
	        \end{scope}
	        \begin{scope}[shift={({0.5 - sqrt(3) / 24}, 0.5)}]
	            \draw[draw = none, fill = yellow!40] (0,0) circle (0.5cm);
	        \end{scope}
	        
    	    \draw[dashed, black!50] (0, 0) grid (\size, 2);
    	    \begin{scope}[shift={({0.5 + sqrt(3) / 6}, {1/2-1/3})}]
        	    \foreach \x in {1, 3, 9, \initiali, 11}
        	    {
        	        \begin{scope}[shift = {(\x, 0)}, scale = \toscale]
                        \import{images/}{blocks.3.blue}
                    \end{scope}
                }
            \end{scope}
    	    \begin{scope}[shift={({0.5 - sqrt(3) / 6}, {1/2-1/3})}]
        	    \foreach \x in {0, 4, 5, 8}
        	    {
        	        \begin{scope}[shift = {(\x, 0)}, scale=\toscale]
                        \import{images/}{blocks.3.red}
                    \end{scope}
                }
            \end{scope}
            
            \begin{scope}[xshift=0.5cm]
                \begin{scope}[yshift=1.75cm]
                    \draw[black!50, dashed] (\ordinali, 0) -- +(0, 1) coordinate (dashline-left);
                    \draw[black!50, dashed] (\initiali, 0) -- +(0, 1) coordinate (dashline-right);
                \end{scope}
                \draw[latex-latex, thick] ($(dashline-left)+(0,-0.25)$) -- ($(dashline-right)+(0,-0.25)$);
                \node[above] at ($(dashline-left)!0.5!(dashline-right)$) {\Large $\left|front(\mathcal{P}_i^{B})\right|$};
                \node at (\ordinali, -0.5) {\Large $i$};
                \node[draw,ellipse,minimum height=2cm,minimum width=5cm] (c) at (4,-3) {\Large
                $front(\mathcal{P}_i^{B})$};
                \draw (-0.2, 0) -- +(0.5, -2) node[below] {\Large $\mathcal{P}^R_{n_1}$};
                \draw (\initiali, 0) -- +(-0.5, -2) node[below] {\Large $\mathcal{P}^B_{i}$};
                \foreach \where/\angle in {0/165, 2/150, 4/90, 5/60, 6/45, 7/30, 8/15}
                {
                    \draw (\where, 0) -- (c.\angle);
                }
            \end{scope}
        \end{scope}
	\end{tikzpicture}}
			\caption{}
			\label{sub:Lemma.5.12.Initial}
		\end{subfigure}
		\par\bigskip
        \begin{subfigure}{\textwidth}
			\resizebox{\textwidth}{!}{    \begin{tikzpicture}
	    \pgfmathsetmacro{\steplength}{1.5}
	    \pgfmathsetmacro{\toscale}{1 / \steplength}
	    
	    \pgfmathsetmacro{\size}{12}
	    \pgfmathtruncatemacro{\ordinali}{4-1}
	    \pgfmathtruncatemacro{\initiali}{10}
	    \pgfmathtruncatemacro{\meetingi}{2}
	    
	    \begin{scope}[scale=\steplength]
	        \draw[draw=none, fill = green!5] (\ordinali, 0) rectangle +(1, 2);
	        \draw[draw=none, fill = green!5!red!5] (\meetingi, 0) rectangle +(1, 2);

	        \begin{scope}[shift={({0.5 + sqrt(3) / 24}, 0.5)}]
	            \draw[draw = none, fill = yellow!40] (\meetingi,1) circle (0.5cm);
	        \end{scope}
	        \begin{scope}[shift={({0.5 - sqrt(3) / 24}, 0.5)}]
	            \draw[draw = none, fill = yellow!40] (\meetingi + 1,0) circle (0.5cm);
	        \end{scope}
	        
    	    \draw[dashed, black!50] (0, 0) grid (\size, 2);
    	    \filldraw[white] (\ordinali+1.5, -0.1) rectangle ++(2, 2.2);
	        \draw[decorate,decoration=snake, black!50]   (\ordinali+1.5, 2)  -- ++(0,-2);
	        \draw[decorate,decoration=snake, black!50]   (\ordinali+1.5+2, 2)  -- ++(0,-2);
            \node at (\ordinali+1.5+1, 1) {\huge $\ldots$};
    	    \begin{scope}[shift={({0.5 + sqrt(3) / 6}, {1/2-1/3})}]
        	    \foreach \x/\y in {0/0, 0/1, 1/1, \meetingi/1, 7/1, 8/0, 9/1}
        	    {
        	        \begin{scope}[shift = {(\x, \y)}, scale = \toscale]
                        \import{images/}{blocks.3.blue}
                    \end{scope}
                }
            \end{scope}
    	    \begin{scope}[shift={({0.5 - sqrt(3) / 6}, {1/2-1/3})}]
        	    \foreach \x in {3, 9, 10, 11}
        	    {
        	        \begin{scope}[shift = {(\x, 0)}, scale=\toscale]
                        \import{images/}{blocks.3.red}
                    \end{scope}
                }
            \end{scope}
            
            \begin{scope}[xshift=0.5cm]
                

                \node at (\ordinali, -0.5) {\Large $i$};
                \node at (\meetingi, -0.5) {\Large $C_i$};
                \draw (\ordinali, 0) -- +(1.2, -0.8) node[below] {\Large $\mathcal{P}^R_{n_1}$};
                \draw (\meetingi, 1) -- +(-1.5, -1.8) node[below] {\Large $\mathcal{P}^B_{i}$};
                \draw (7, 1.2) -- +(-0.5, -2);
                \draw (8, 0.2) -- +(0, -1);
                \draw (9, 1.2) -- +(1, -2.) node[label={[shift={(-2.5, -1.2)}]\Large $\left|back(\mathcal{P}_i^B)\right|$ unmet agents}]{};
            \end{scope}
        \end{scope}
	\end{tikzpicture}}
			\caption{}
			\label{sub:Lemma.5.12.Case.1}
		\end{subfigure}
		\par\bigskip
        \begin{subfigure}{\textwidth}
			\resizebox{\textwidth}{!}{    \begin{tikzpicture}
	    \pgfmathsetmacro{\steplength}{1.5}
	    \pgfmathsetmacro{\toscale}{1 / \steplength}

	    \pgfmathsetmacro{\size}{12}
	    \pgfmathtruncatemacro{\ordinali}{1}
	    \pgfmathtruncatemacro{\initiali}{10}
	    \pgfmathtruncatemacro{\meetingi}{6}
	    
	    \begin{scope}[scale=\steplength]
	        \draw[draw=none, fill = green!5] (\ordinali, 0) rectangle +(1, 2);
	        \draw[draw=none, fill = green!5!red!5] (\meetingi, 0) rectangle +(1, 2);
	        \draw[draw=none, fill = blue!5] (\initiali, 0) rectangle +(1, 2);

	        \begin{scope}[shift={({0.5 + sqrt(3) / 24}, 0.5)}]
	            \draw[draw = none, fill = yellow!40] (\meetingi,0) circle (0.5cm);
	        \end{scope}
	        \begin{scope}[shift={({0.5 - sqrt(3) / 24}, 0.5)}]
	            \draw[draw = none, fill = yellow!40] (\meetingi,1) circle (0.5cm);
	        \end{scope}
	        
    	    \draw[dashed, black!50] (0, 0) grid (\size, 2);
    	    \foreach \x in { 3.5, 8.5}
    	    {
    	        \begin{scope}[xshift=\x cm]
            	    \filldraw[white] (0, -0.1) rectangle ++(1, 2.2);
        	        \draw[decorate,decoration=snake, black!50]   (0, 2)  -- ++(0,-2);
        	        \draw[decorate,decoration=snake, black!50]   (1, 2)  -- ++(0,-2);
    	            \node at (0.5, 1) {\huge $\ldots$};
	            \end{scope}
	        }
	        \filldraw[white] (9.3, -0.1) rectangle (\size, 2.1);
	        
    	    \begin{scope}[shift={({0.5 + sqrt(3) / 6}, {1/2-1/3})}]
    	        \foreach \blue in {0, 1}
    	        {
        	        \begin{scope}[shift = {(\meetingi + \blue, \blue)}, scale = \toscale]
                        \import{images/}{blocks.3.blue}
                    \end{scope}
                }
            \end{scope}
    	    \begin{scope}[shift={({0.5 - sqrt(3) / 6}, {1/2-1/3})}]
    	        \begin{scope}[shift = {(\meetingi, 1)}, scale=\toscale]
                    \import{images/}{blocks.3.red}
                \end{scope}
            \end{scope}
            
            \begin{scope}[xshift=0.5cm]
                \node[right] at (\ordinali, -0.5) {\Large $i$};
                \node[right] at (\ordinali + 1, -0.5) {\Large $i + 1$};
                \node[left] at (\meetingi, -0.5) {\Large $C_i$};
                \draw (\meetingi+0.25, 0) -- +(1, -.45) node[right] {\Large $\mathcal{P}^B_{i}$};
                \draw (\meetingi+1, 1) -- +(2+0.25, -.45-1) node[right] {\Large $\mathcal{P}^B_{i+1}$};
                \draw (\meetingi-0.25, 1.05) -- +(-1.25, -1.45) node[left] {\Large $\mathcal{P}^R_{n_1}$};
            \end{scope}
        \end{scope}
	\end{tikzpicture}}
			\caption{}
			\label{sub:Lemma.5.12.Case.2}
		\end{subfigure}
    \end{minipage}
    \caption{\cref{lemma:lower.bound}. (\subref{sub:Lemma.5.12.Initial}) At time $t=0$, $\mathcal{P}^B_i$ is at distance $|front(\mathcal{P})|$ from column $i$. (\subref{sub:Lemma.5.12.Case.1}) Case 1: $C_i \leq i$. $\mathcal{P}^R_{n_1}$ must meet at least $|back(\mathcal{P}^i_B)|$ blue agents after meeting $\mathcal{P}^i_B$.  (\subref{sub:Lemma.5.12.Case.2}) Case 2: $C_i > i$. $\mathcal{P}^B_{i+1}$ needs at least two time steps to enter column $C_i$, and at least $b_i+1$ time steps to arrive at column $i+1$.}
\end{figure}

\begin{lemma}\label{lemma:t_icrudelowerboundlemma}
    Let $t_i$ be the first time when $\mathcal{P}^B_i$ reaches column $i$, i.e. an agent with label $i$ arrives at column $i$ for the first time. Then $t_i \geq |front(\mathcal{P}^B_i)| + \mathcal{V}_i^{\textit{B}}$.
\end{lemma}
\begin{proof}
Since in the initial configuration $\mathcal{P}^B_i$ sees $i-1$ blue agents in front of it, we have that
\begin{equation*}
\left|(\mathcal{P}^B_i)_x - i\right| = \left|front(\mathcal{P}^B_i)\right|
\end{equation*} 
(e.g., see \cref{sub:Lemma.5.12.Initial}). Hence $\mathcal{P}^B_i$ requires at least  $|front(\mathcal{P}^B_i)|$ horizontal moves to get to column $i$, and hence requires this many time steps (\cref{observation:TagSlowMovement}). 
If $\mathcal{V}_i^{\textit{B}} = 1$ then ${\mathcal{P}^B_i}$ performs a non-leftward movement before arriving at column $i$, and this adds one time step.
\end{proof}

\begin{observation}
    \label{observation:labelsmeetoneatatime}
    A given label $P^B$ cannot meet more than one label at any time step.
\end{observation}

\begin{proof}
     A given agent $A$ cannot meet more than one agent at any time step, since there are only two rows in the grid (see \cref{fig:example.configs}). Due to \cref{observation:TagSlowMovement}, the same is true of labels.
\end{proof}

We now proceed to the proof of the main technical lemma of this section.

\begin{lemma}\label{lemma:lower.bound}
    Let $t_{final}$ be the makespan of \textit{ALG} on the given initial configuration. Then for any $1 \leq i \leq n_2$,
    \begin{align*}
        t_{final} \geq \left|front\left(\mathcal{P}^B_i\left(0\right)\right)\right|  + \left|back\left(\mathcal{P}^B_i\left(0\right)\right)\right| + \mathcal{V}_i^{\textit{B}} 
    \end{align*}
\end{lemma}
\begin{proof}
At time $t_{final}$, $P^B_i$ must be located at or to the left of column $i$ (see \cref{sub:Lemma.5.12.Initial}). Let $t_i$ denote the first time $P^B_i$ reaches column $i$, and let  $$b_i = |front(\mathcal{P}^B_i)| + \mathcal{V}_i^{\textit{B}}.$$ In \cref{lemma:t_icrudelowerboundlemma} we show that $t_i \geq b_i$. Note that $t_i = b_i$ if and only if $\mathcal{P}^B_i$ moves left at all times $t < t_i$ except at most $\mathcal{V}_i^{\textit{B}}$ time steps. 

We will prove the lemma by tracking $\mathcal{P}^R_{n_1}$. At time $t=0$, since the initial configuration is normal, $\mathcal{P}^R_{n_1}$ is the leftmost agent. Hence, between time $0$ and $t_{final}$ it ought meet $\mathcal{P}^B_i$. Let $M_i$ be the time of their first meeting, and $C_i$ be the $x$-coordinate of $\mathcal{P}^B_i$ at time $M_i$. 

We separate the proof into cases. 

\textit{Case 1.}  Assume $C_i \leq i$. If $C_i \leq i$ then $M_i \geq t_i \geq b_i$. Since in a sorted configuration $\mathcal{P}^R_{n_1}$ must be located to the right of all blue-occupied and empty columns, $\mathcal{P}^R_{n_1}$ must meet the blue labels $\mathcal{P}^B_{i+1}, \mathcal{P}^B_{i+2}, \ldots, \mathcal{P}^B_{n_2}$ before time $t_{final}$. By \cref{observation:labelsmeetoneatatime}, this must take 
$n_2 - i = |back(\mathcal{P}^B_i)|$ time ticks. Hence in total \textit{ALG} requires at least $|front(\mathcal{P}^B_i)| + |back(\mathcal{P}^B_i)|+ \mathcal{V}_i^{\textit{B}} $ time steps to complete, as desired. (See \cref{sub:Lemma.5.12.Case.1})

\textit{Case 2.} Assume $C_i > i$. In this case, we  claim that $\mathcal{P}^B_{i+1}$ cannot reach column $C_i$ before time $M_i + 2$. If column $C_i$ is a mixed column at time $M_i$ then no agent can move into $C_i$ at time $M_i+1$ (see \cref{sub:Lemma.5.12.Case.2}). Otherwise, $C_i$ is occupied by $\mathcal{P}^B_i$, while $\mathcal{P}^R_{n_1}$ occupies $C_i+1$ in a different row than $\mathcal{P}^B_i$. Since both rows are obstructed by an agent, there is no empty location that will enable $\mathcal{P}^B_{i+1}$ to move into column $C_i$ before time $M_i+2$.

The label $\mathcal{P}^B_{i+1}$ must reach column $i+1$ before time $t_{final}$. By the above, $\mathcal{P}^B_{i+1}$ can reach column $C_i$ for the first time only $2$ time steps after $\mathcal{P}^B_{i}$. The fastest $\mathcal{P}^B_{i}$ can reach column $i+1$ is within $b_i - 1$ time steps, and this can only occur when it is able to move left at every time step after it reaches column $C_i$. Hence the fastest time $\mathcal{P}^B_{i+1}$ can reach column $i+1$ is $b_i - 1 + 2$. 

We can now repeat the argument above for $\mathcal{P}^B_{i+1}$. Note that by \cref{observation:labelsmeetoneatatime}, $\mathcal{P}^B_{i+1}$ cannot meet $\mathcal{P}^R_{n_1}$ before time $M_i+1$. We split the argument into the same two cases: 

In Case 1, we assume $\mathcal{P}^B_{i+1}$ reaches column $i+1$ before it meets  $\mathcal{P}^R_{n_1}$ or at the same time step. Moreover, $\mathcal{P}^R_{n_1}$ must spend $n_2 - i - 1 = |back(\mathcal{P}^B_i)| - 1$ time steps meeting the blue labels $\mathcal{P}^B_{i+2}, \mathcal{P}^B_{i+3}, \ldots, \mathcal{P}^B_{n_2}$. Since $\mathcal{P}^B_{i+1}$ needs at least $b_i - 1 + 2$ time steps to reach column $i+1$, we have that $t_{final} \geq b_i + |back(\mathcal{P}^B_i)|$ just like before, and we are done.

In Case 2, $\mathcal{P}^B_{i+1}$ meets $\mathcal{P}^R_{n_1}$ before reaching column $i+1$. Denote the $x$-coordinate of $\mathcal{P}^B_{i+1}$  at the time of the meeting by $C_{i+1}$. By the same argument as before, $\mathcal{P}^B_{i+2}$ can arrive at column $C_{i+1}$ only $2$ time steps after $\mathcal{P}^B_{i+1}$. $\mathcal{P}^B_{i+1}$ can arrive at column $i+2$ within $b_i+1-1$ time steps only if it moves left at every time step after meeting $\mathcal{P}^R_{n_1}$. Hence like in the previous argument, the fastest time $\mathcal{P}^B_{i+2}$ can reach column $i+2$ is $b_i + 2$. 

We can extend this exact argument by induction to all the blue agents $\mathcal{P}^B_{i+3}$, $\ldots$, $\mathcal{P}^B_{n_2}$. At every stage of the induction we separate into two cases. In the first case, $\mathcal{P}^B_{j}$ arrives at column $j$ before it meets $\mathcal{P}^R_{n_1}$ or at the same time step, in which case we can show that $t_{final} \geq b_i + |back(\mathcal{P}^B_i)|$. In the second case, $\mathcal{P}^B_{j}$ meets $\mathcal{P}^R_{n_1}$ before reaching column $j$, which implies it needs at least $b_i + (j - i)$ time steps to arrive at column $j$. By continuing the induction up to $\mathcal{P}^B_{n_2}$ we deduce that the makespan must be at least $$b_i + (n_2 - i) = b_i + |back(\mathcal{P}^B_i)| = |front(\mathcal{P}^B_i)|  + |back(\mathcal{P}^B_i)| + \mathcal{V}_i^{B} $$ in all cases, so we are done. 
\end{proof}

\cref{lemma:lower.bound} provides a lower bound on the makespan of any sorting algorithm $ALG$ with respect to a given initial configuration. The lower bound is based on the $front$ and $back$s of the blue agents and on the $\mathcal{V}_i^{B}$ values induced by $ALG$. By symmetry, a similar argument holds for the red agents, and consequently we may establish:

\begin{corollary}\label{lemma:optimal-lower-bound}
    Let $t_{final}$ be the makespan of \textit{ALG} on the given initial configuration. Then
    \begin{equation}\label{equation_optimal_lower_bound}
        t_{final} \geq \max\bigg(\max\limits_{1 \leq i \leq n_1} \big(f(P_i^R(0)) + \mathcal{V}_i^{R}\big),\max\limits_{1 \leq i \leq n_2} \big(f(P_i^B(0)) + \mathcal{V}_i^{B}\big)\bigg)
    \end{equation} 
\end{corollary}

The somewhat unwieldy expression in \cref{lemma:optimal-lower-bound} just says that $t_{final}$ is bounded below by $f_{max}$ plus at most $1$, depending on the value of the $\mathcal{V}_i$s. Recalling \cref{Vmaxdef}, we would like to show that the right-hand side of  \eqref{equation_optimal_lower_bound} is at least as large as $f_{max} + \mathcal{V}$. To show this will require a bit more work. 

\begin{lemma}\label{lemma:critical-agents-face-each-other}
If a normal initial configuration of agents has both red and blue critical agents, then there must be some blue critical agent to the right of some red critical agent. 
\end{lemma}

\begin{figure}[!ht]
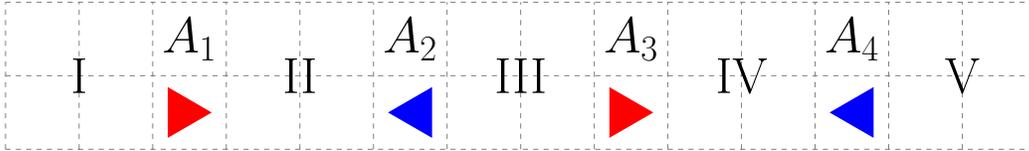

    \centering
	\resizebox{\textwidth}{!}{	\begin{tikzpicture}
	    \pgfmathsetmacro{\steplength}{1.5}
	    \begin{scope}[scale=\steplength]
    	    \draw[dashed, black!50] (0, 0) grid (14, 2);
    	    \begin{scope}[shift={({-0.5 + sqrt(3) / 6}, {1/2-1/3})}]
        	    \foreach \x in {6, 12}
        	    {
        	        \begin{scope}[xshift = \x cm, scale=1/\steplength]
                        \import{images/}{blocks.3.blue}
                    \end{scope}
                }
            \end{scope}
    	    \begin{scope}[shift={({0.5 - sqrt(3) / 6}, {1/2-1/3})}]
        	    \foreach \x in {2, 8}
        	    {
        	        \begin{scope}[xshift = \x cm, scale=2/3]
                        \import{images/}{blocks.3.red}
                    \end{scope}
                }
            \end{scope}
            \begin{scope}[shift={(0.5, 1.5 )}]
                \foreach \i in {1, ..., 4}
                {
                    \node at ({-1 + 3 * \i}, 0) {\huge $A_\i$};
                }
            \end{scope}
            \begin{scope}[shift={(0, 1 )}]
                \foreach \i/\name in {1/I, 2/II, 3/III, 4/IV, 5/V}
                {
                    \node at ({-2 + 3 * \i}, 0) {\huge \name};
                }
            \end{scope}
        \end{scope}
	\end{tikzpicture}}
    \caption{A fictitious configuration with critical agents $A_2$ (blue) and $A_3$ (red) not facing each other, per the construction given in \cref{lemma:critical-agents-face-each-other}. Column ranges between agents are marked ($I-V$).}
    \label{fig:general.config.faceless.critical.agents}
\end{figure}
\begin{proof}
    To prove the Lemma we will assume that the claim is wrong and deduce a contradiction. 
    
    Suppose there exists an initial configuration with red and blue critical agents, where all the red critical agents are to the right of all the blue critical agents. Let $A_2$ be some blue critical agent and $A_3$ be some red critical agent. Let $A_1$ be the right-most red agent to the left of $A_2$, and $A_4$ the left-most blue agent to the right of $A_3$. We know these agents exist since the initial configuration is normal. Denote the areas between the agents in the following manner: $I$ denotes the set of columns to the left of $A_1$, $II$ denotes the columns between $A_1$ and $A_2$, and so forth until $V$, which denotes the columns to the right of $A_4$. This construction is illustrated in \cref{fig:general.config.faceless.critical.agents}.
    
    Denote by $\mathbf{E}_x, \mathbf{R}_x$ and $\mathbf{B}_x$, respectively, the number of empty, red-occupied, and blue-occupied columns in the set $x \in \{I,II,\ldots, V\}$.  We will write down the value of $f(\cdot)$ for all four agents $A_i$. By definition,
    
    \begin{equation*}
        f(A_2) = \mathbf{R}_{I} + \mathbf{E}_{I} + 1 + \mathbf{E}_{II} + \mathbf{B}_{III} + 1 + \mathbf{B}_{V},
    \end{equation*}
    where the first $+1$ counts $A_1$, and the second $+1$ counts $A_4$.
    Similarly,
    \begin{equation*}
        f(A_3) = \mathbf{R}_{I} + 1 + \mathbf{R}_{III} + \mathbf{E}_{IV} + 1 + \mathbf{B}_{V}+ \mathbf{E}_{V}
    \end{equation*}
    \begin{equation*}
        f(A_1) = \mathbf{R}_{I} + \mathbf{B}_{II} + \mathbf{E}_{II} + 1 + \mathbf{B}_{III} + \mathbf{E}_{III} + \mathbf{E}_{IV} + 1 + \mathbf{B}_{V} + \mathbf{E}_{V}
    \end{equation*}
    \begin{equation*}
        f(A_4) = \mathbf{R}_{I} + \mathbf{E}_{I} + 1 + \mathbf{E}_{II} + \mathbf{R}_{III} + \mathbf{E}_{III} + 1 + \mathbf{R}_{IV} + \mathbf{E}_{IV} + \mathbf{B}_{V}
    \end{equation*}
    
    By algebra, we have that if $\mathbf{B}_{III} \geq \mathbf{R}_{III}$ then $f(A_1) \geq f(A_3)$, and otherwise $f(A_4) > f(A_2)$. In the case $f(A_1) \geq f(A_3)$, we have by definition that $A_1$ is a red critical agent to the left of the blue critical agent $A_2$, in contradiction to our initial assumption. In the case $f(A_4) > f(A_2)$, we arrive at a contradiction, since $f(A_4)$ cannot exceed $f_{max} = f(A_2)$. This completes the proof.
\end{proof}

Let us now show that \cref{lemma:optimal-lower-bound} bounds from above our desired lower bound of $f_{max} + \mathcal{V}$, by proving the following:

\begin{lemma}
\label{lemma:inequality_v_final}
\begin{equation}
\label{equation:inequality_v_final}
    \max\bigg(\max\limits_{1 \leq i \leq n_1} \big(f(P_i^R(0)) + \mathcal{V}_i^{R}\big),\max\limits_{1 \leq i \leq n_2} \big(f(P_i^B(0)) + \mathcal{V}_i^{B}\big)\bigg) \geq f_{max} + \mathcal{V}
\end{equation}
\end{lemma}

\begin{proof}
The case where $\mathcal{V} = 0$ is trivial, since the maximum of $f$ taken over all agents is by definition $f_{max}$ and our  $\mathcal{V}_i^R, \mathcal{V}_i^B$s are non-negative. So let us assume $\mathcal{V} = 1$. By definition, this means either (a) there is a critical agent $A$ that sees another agent immediately in front of it in the initial configuration, or (b) there are two critical agents of different colors in the initial configuration.

In case (a), let us suppose without loss of generality that $A$ is red, and let $ord(A) = i$. Then necessarily $f(P_i^R(0)) = f_{max}$ and, since $A$ cannot move horizontally nor change labels in the first time step of $ALG$, $\mathcal{V}_i^R = 1$. Hence $f(P_i^R(0)) + \mathcal{V}_i^R = f_{max} + \mathcal{V}$ which proves Inequality \eqref{equation:inequality_v_final}.

In case (b), \cref{lemma:critical-agents-face-each-other} tells us that there exists a pair of \textit{red and blue critical agents} $A^R$ and $A^B$ facing each other. Let $ord(A^R) = i$ and $ord(A^B) = j$. If $A^R$ and $A^B$ do not change labels before meeting (i.e., for all times $t$ prior to the meeting, $ord(A^R,t) = i$ and $ord(A^B,t) = j$), then necessarily one of the labels $\mathcal{P}_i^R$ or $\mathcal{P}_j^B$ must move to the upper row so that they can bypass each other. Hence $\mathcal{V}_i^R = 1$ or $\mathcal{V}_j^B = 1$, and similar to case (a), this establishes Inequality \eqref{equation:inequality_v_final}.

Otherwise, either $A^R$ or $A^B$ changes their label before meeting. Let us suppose wlog that $A^R$ changes its label at time $t_0$, i.e. $ord(A^R, t_0-1) = i \land ord(A^R, t_0) \neq i$. At $t_0$, $A^R$'s label can be either $i+1$ or $i-1$. It is straightforward to verify by hand that in either case, $\mathcal{V}^R_i = 1$, since in order for two adjacent agents to swap labels, one of them must wait or move toward the other, and this causes $\mathcal{V}^R$ to become $1$ for both agents that swapped labels. Hence, similar to case (a) and (b), this establishes Inequality \eqref{equation:inequality_v_final}. 
\end{proof}

\cref{lemma:optimal-lower-bound} and \cref{lemma:inequality_v_final} together establish the lower bound $f_{max}+\mathcal{V}$ of \cref{theorem:makespanresult}. It remains to show an algorithm  that achieves this bound exactly. 

\section{Optimal algorithm for normal configurations}
\label{sec:TheAlgorithm}
In this section we present a simple agent sorting algorithm. We show that despite its simplicity, the algorithm is optimal for normal initial configurations, in the sense that its makespan \textbf{always} meets the lower bound established in  \cref{theorem:makespanresult}, thus completing the proof of \cref{theorem:makespanresult}.

	\begin{algorithm}
		\SetAlgoLined
		\KwResult{The configuration is sorted. }
		\kwInit{$t \gets 0$}

		\ForEach{agent  $A \in \mathcal{A}$}
		{
		    calculate $f(A)$ to determine the critical agents\;
		}
		\uIf{there are critical agents of both colors}{
		    $c \gets \text{red}$\;
		}
		\Else{
		    $c \gets$ the color of critical agents\;
		}
		\ForEach{agent $A \in \mathcal{A}$}{
		    \uIf{$A$'s color is $c$}{
		        move one step in the desired direction if the adjacent location in that direction is not occupied at the beginning of time  $t=0$
		    }
		    \Else{
		        move to the second row\;
		    }
		}
		\While{configuration is not sorted} {
			$t \gets t + 1$\;
			\ForEach{red agent $\in \mathcal{A}$}{
				\If{right-adjacent location is empty at the beginning of $t$} {
					move one step right\;
				}
			}
			\ForEach{blue agent $\in \mathcal{A}$}{
				\If{left-adjacent location is empty at the beginning of $t$} {
					move one step left\;
				}
			}
		}
		\caption{Optimal sorting algorithm for unsorted normal initial configurations.}
		\label{algo:trivial.centralized}
	\end{algorithm}		

    \cref{algo:trivial.centralized} is simple to describe: \textit{aim to move in your desired direction},
    left for blue agents and right for red agents. The only twist is the first tick, where either all red or all blue agents move to the second row. Which color should we move vertically? The answer comes from \cref{theorem:makespanresult}; we want to avoid paying more than $f_{max} + \mathcal{V}$ time steps by moving vertically only the color without critical agents.  This enables critical agents to move horizontally without being delayed by a vertical movement. If there are red critical agents - we move blue agents to the second row, otherwise red agents are moved.
    
    Denote the above algorithm ${ALG}_1$. To establish its makespan, let us first study the simple scenario where all blue agents are in the one row and all red agents are in the other, and at every time step they simply move horizontally in the desired direction:
    
    \begin{lemma}\label{theorem:strange-model-algo-runtime} 
        Consider a normal configuration where all red agents are at the top row and all blue agents are at the bottom row, and every column contains at most one agent.
        
        Let $ALG^{\bot}_1$ denote the algorithm that says: at every time step, red agents move right unless the adjacent location to their right is occupied, and blue agents move left unless the adjacent location to their left is occupied. Then the makespan of $ALG^{\bot}_1$ on this configuration is at most $f_{max}$.
    \end{lemma}
    
    \begin{proof}
        For any red agent $R$, we define $front^{\bot}(R, t)$ as the set of all empty locations on $R$'s row which are to the right of $R$ at time $t$. Furthermore, define $back^{\bot}(R,t)$ to be the set of all red agents to the left of $R$. 
        
        Similarly, for any blue agent $B$, we define $front^{\bot}(B, t)$ as the set of all empty locations on $B$'s row which are to the left of $B$ at time $t$, and define $back^{\bot}(B,t)$ to be the set of all blue agents to the right of $B$.
        
        For every agent $A$ define $f^{\bot}(A,t)$ at time $t$ in the following manner:
        \begin{itemize}
            \item if $front^{\bot}(A,t)$ contains no empty locations, $f^{\bot}(A,t)=0$
            \item otherwise, $f^{\bot}(A,t) = |front^{\bot}(A,t)|+|back^{\bot}(A,t)|$
        \end{itemize}
        
       In the Lemma's assumed agent configuration, $|front^{\bot}(A,0)| > 0$ for any agent $A$. Hence for all agents $f^{\bot}(A,0) \neq 0$. Recalling \cref{def:f}, this means $f^{\bot}(A,0)=f(A)$ for any agent $A$. Hence, there is a critical agent $A$ for which $f^{\bot}(A,0)=f_{max}$. 
       
       Let us define $f_{max}^{\bot}(t) = \max_{A \in \mathcal{A}}f^{\bot}(A,t)$. By the above, $f_{max}^{\bot}(0)=f_{max}$. When $f_{max}^{\bot}(t) = 0$, the agent configuration is sorted, since every agent has moved as far as it could in its desired direction.  We will show that $f_{max}^{\bot}(t)$ decreases every time step as long as it is not $0$, which completes the proof.
       
       At time $t$, let $A^*$ be an agent for which $f^{\bot}(A^*,t)=f_{max}^{\bot}(t)$. Let us suppose w.l.o.g. that $A^*$ is a red agent. Note that if a red agent $R$ is an adjacent rightward neighbor of $A^*$ at time $t$ then 
       \begin{enumerate}[label=(\alph*)] 
            \item if $f^{\bot}(R, t) = 0$ then $f^{\bot}(A^*, t) = 0$, and
            \item \label{enum:option-b} $f^{\bot}(A^*,t) = f^{\bot}(R, t) - 1$ otherwise.
        \end{enumerate} 
        
        \ref{enum:option-b} is of course impossible, since $A^*$ maximizes $f^{\bot}(\cdot,t)$. Hence, if $A^*$ cannot move right at time $t$, we have that $f^{\bot}(A^*,t) = f_{max}(t) = 0$, so the configuration is sorted. Otherwise, $A^*$ sees an empty location immediately to its right, and it moves to it at time $t$. This causes $f^{\bot}(A^*,t)$ to decrease by $1$.
        
        Since the above argument is true for \textit{any} agent that maximizes $f^{\bot}(\cdot, t)$, we see that as long as $f_{max}^{\bot}(t) \neq 0$, $f_{max}$ decreases at every time step.
    \end{proof}
    
    An immediate corollary of \cref{theorem:strange-model-algo-runtime}, which (as mentioned in the introduction) relates to asymmetric simple exclusion processes and may be of independent interest, is:

    \begin{corollary}
    \label{corollary:singlerowtime}
    Assume $j$ agents located on a single row (without access to a spare row) move right at every time step where they are unobstructed by another agent. Then after exactly $f_{max}$ time steps,
    the $j$ right-most columns are occupied by an agent.
    \end{corollary}

    
    \begin{theorem}
    \label{theorem:optimal-sorting-alg1}
        For a given initial normal configuration, ${ALG}_1$'s makespan is $f_{max} + \mathcal{V}$.
    \end{theorem}
    
    \begin{proof}
        By \cref{theorem:makespanresult}, we know the makespan is at least  $f_{max} + \mathcal{V}$. Let us show it is at most this.
        
        Let us define $f_{max}'$ to be calculated like $f_{max}$ in \cref{def:f}, but over the agent configuration at time $t=1$. Note that $f_{max} \geq f_{max}'$, since $ALG_1$ never moves agents in a way that can increase this value. 
        
        Note that from time $t=1$ onwards, all agents are on two separate rows and execute $ALG_1^{\bot}$ (of \cref{theorem:strange-model-algo-runtime}) in their respective row. Hence, by \cref{theorem:strange-model-algo-runtime}, our algorithm's makespan is at most $f_{max}' + 1 \leq f_{max} + 1$. Hence, if $\mathcal{V} = 1$, we are done.
        
        Let us deal with the case $\mathcal{V} = 0$. We assume, w.l.o.g., that there are red critical agents. Since $\mathcal{V} = 0$, this means there are no blue critical agents, and also that every red critical agent takes a step to the right in the first time step of the algorithm. Since the maximum value of $f$ is obtained over a red agent, we know that
        
        \begin{equation*}
            \max\limits_{B \in \mathcal{B}} f(B) \leq f_{max} - 1.
        \end{equation*}
        
        Furthermore, since every red critical agent moves right at time $t=0$, when we re-calculate $f$ over the configuration at time $t=1$, $f_{max}$ will have decreased by $1$ (since every critical agent $A$ has shifted an empty location from $front(A)$ to $back(A)$). Hence $f_{max}' = f_{max} - 1$, and so our makespan is at most $f_{max}$, as claimed. 
    \end{proof}
    
    \cref{theorem:optimal-sorting-alg1} and the lower bound result of the previous section establish \cref{theorem:makespanresult}. We see that ${ALG}_1$ is an optimal algorithm for physically sorting any normal initial configuration. In the next section, we will extend this result to non-normal configurations.

\section{Non-normal initial configurations}
\label{NonNormalConfigurationSection}
    In the previous sections we have handled a subset of all the possible initial configurations - the normal configurations (\cref{def:normal-configuration}). In this section, we extend our previously derived lower bound to all possible initial configurations, and extend our optimal algorithm to such configurations.

    \begin{definition}
       Consider the configuration of agents at time $t$. Let $R$ be the leftmost red agent and let $B$ be the rightmost blue agent. The \textbf{maximal normal sub-configuration at time $t$} is the set of all columns in the interval $[R_x(t), B_x(t)]$. Note that this set is empty when $R_x(t) > B_x(t)$.
       
       Furthermore, let $\mathcal{S}$ be the (possibly empty) maximal normal sub-configuration at time $t=0$. Let $\mathcal{S}^c = \mathcal{A} \setminus \mathcal{S}$ be the set of columns outside $\mathcal{S}$.
    \end{definition}
    
    We define a function $f^*$ to equal $f$ for all agents in (the columns of) $\mathcal{S}$, and for any agent $A$ in $\mathcal{S}^c$ we set $f^*(A) = |front(A)|$. We further define $f^*_{max} = \max\limits_{A \in \mathcal{A}} f^*(A)$.
    
    \begin{definition}\label{def:f*-critical agents}
        An agent $A \in \mathcal{A}$ for which $f^*(A) = f^*_{max}$ is called an \textbf{$f^*$-critical agent}.
    \end{definition}
    
    We will define $\mathcal{V}^*$ to closely resemble $\mathcal{V}$:
    \begin{definition}\label{def:V*}
        $\mathcal{V}^*$ will equal $1$ if:
        \begin{enumerate}[label={\arabic*.}]
            \item There is both a red and a blue $f^*$-critical agent \textbf{and} $\mathcal{S} \neq \varnothing$, or
            \item There is an $f^*$-critical agent such that, in the initial configuration, another agent is located immediately in front of it \textbf{and} $f^*_{max} > 0$.
        \end{enumerate}
        Otherwise, $\mathcal{V}^* = 0$.
    \end{definition}
    
    \cref{lemma:lower.bound} can now be extended to non-normal initial configurations as follows:
        
    \begin{lemma}
    \label{lemma:boundextended}
        The makespan of any algorithm that brings the initial configuration to a sorted configuration is at least $f_{max}^* + \mathcal{V}^*$.
    \end{lemma}

    \begin{proof}
        The argument is a rather straightforward generalization of the previous sections.
    
        We assume first that $\mathcal{S} \neq \varnothing$. We note that in this case, agents outside $\mathcal{S}$ are not $f^*$-critical. Indeed, for any same-colored agents $A_1 \in \mathcal{S}$ and $A_2 \notin \mathcal{S}$, $front(A_2) \subsetneq front(A_1)$. Therefore, we trivially have that $f^*(A_2) < f^*(A_1)$. We note that the argument in \cref{lemma:lower.bound} generalizes genuinely to agents inside $\mathcal{S}$. Consequently, \cref{lemma:lower.bound} provides a lower bound on makespan, since $f \equiv f^*$ on $\mathcal{S}$ and when restricted to this set, $\mathcal{V}^*$ is equivalent to the definition of $\mathcal{V}$.
        
        Now let us assume $\mathcal{S} = \varnothing$, i.e. in the initial configuration no pair of red and blue agents face each other. Each label $\mathcal{P}^B_i$ must traverse at least $|front(\mathcal{P}^B_i(0))|$ columns before a sorted configuration can be reached and the same claim is true for $\mathcal{P}^R_i$. In particular, by definition, $f^*$-critical agents will need to traverse $f^*_{max}$ columns to reach their position. If $\mathcal{V}^* = 0$, this establishes our desired lower bound.
        
        If $\mathcal{V}^* = 1$, we have that there is an $f^*$-critical agent that is blocked by an adjacent agent, hence cannot change columns in the first time step. Furthermore, since $f^*_{max} > 0$, the configuration is not sorted, so this agent \textit{must} traverse at least one column. Hence, at least one $f^*$-critical agent must spend $\mathcal{V}^*$ ticks before starting to move horizontally, establishing a lower bound of $f^*_{max} + \mathcal{V}^*$ time steps before a sorted configuration is reached.
    \end{proof}

    We are now interested in designing an optimal physical sorting algorithm for non-normal configurations -  \cref{algo:general.movement}. The algorithm we present will work as follows: the agents that are inside $S$ at time $t=0$ will continue to execute \cref{algo:trivial.centralized} as before, whereas all agents initialized outside $\mathcal{S}$ will execute an ``alternating row split'' strategy. These strategies are executed independently by the two sets of agents: agents initialized in $\mathcal{S}$ will not reach the $\mathcal{S}^c$ columns fast enough to interact with agents initialized outside $\mathcal{S}^c$.

    \begin{definition}
        Let $r_0$ be the label of the leftmost red agent in $\mathcal{S}^c$ and $b_0$ be the label of the rightmost blue agent in $\mathcal{S}^c$ at time $t=0$. $\mathcal{P}^R_{r_0}(0)$ and $\mathcal{P}^B_{b_0}(0)$ will be called red and blue \textbf{anchor} agents respectively.
        \label{definition:anchoragents}
    \end{definition}
	
	\begin{algorithm}
	    \footnotesize
		\SetAlgoLined
		\KwResult{The configuration is sorted.}
	    \kwInit{$t \gets 0$, $\mathcal{Q} \gets$ all agents initialized in the columns of $\mathcal{S}$, $\mathcal{P} \gets$ all agents initialized in the columns of $\mathcal{S}^c$}
				
		\ForEach{agent $A \in \mathcal{A}$}
		{
		    calculate $f^*(A)$ to determine the critical agents\;
		}
		\uIf{there are $f^*$-critical agents of both colors}{
		    $c \gets \text{red}$\;
		}
		\Else{
		    $c \gets$ the color of the $f^*$-critical agents\;
		}
		move all agents in $\mathcal{Q}$ whose color is not $c$ to the second row\;
		
		\ForEach(\tcp*[h]{agents initialized in $S^c$ split themselves between the upper and lower rows}){agent $A \in \mathcal{P}$}{
       		\uIf{$A$ is red \And ($r_0 - ord(A, 0)$) is odd}{
    		    move $A$ to the second row\;
    		}
    		\ElseIf{$A$ is blue \And ($b_0 - ord(A, 0)$) is odd}{
    		    move $A$ to the second row\;
    		}
    	}
    	
		\ForEach{agent $A \in \mathcal{A}$}{
		    \If{$A$ has not moved due to a previous step of the algorithm}{
		        move one step in the desired direction (left for blue agents, right for red agents) if the adjacent location in that direction is not occupied at the beginning of time $t=0$\;
		    }
		}
    	
    	\While(\tcp*[h]{at times $t\geq1$ all agents move horizontally toward their destination}){the configuration is not sorted} {
			$t \gets t + 1$\;
			\ForEach{red agent $A$}{
				\If{there is a mixed, empty, or blue-occupied column to the right of $A$  at the beginning of time $t$} {
					move one step right\;
				}
			}
			\ForEach{blue agent $A$}{
				\If{there is a mixed, empty, or red-occupied column to the left of $B$ at the beginning of time $t$} {
					move one step left\;
				}
			}
		}
		\caption{Optimal sorting algorithm for any unsorted initial configuration.}
	    \label{algo:general.movement}
    \end{algorithm}

    The ``alternating row split'' strategy is based on the following simple idea: at time $t=0$, split agents outside $\mathcal{S}$ in an alternating fashion between the two rows, so that there is at least one empty space between each pair of same-colored agents in the same row (see \cref{figure:subnormalalgoexample}). Once this is done, each agent in $\mathcal{S}^c$ can move horizontally in its desired direction at every time step without any further delays - see \cref{figure:subnormalalgoexample}. 
    
    \begin{figure}[!ht]
    \centering
    \foreach \tick in {0, ..., 4}
    {
    \begin{subfigure}{\textwidth}
		\resizebox{\textwidth}{!}{	\begin{tikzpicture}
	    \pgfmathsetmacro{\steplength}{1.5}
	    \pgfmathsetmacro{\toscale}{2/3}
	    \pgfmathsetmacro{\size}{25}
	    \begin{scope}[scale=\steplength]
    	    \ifthenelse{\tick=0}
    	    {
    	        \draw[draw=none, fill = blue!5] (0, 0) rectangle (9, 2);
    	        \draw[draw=none, fill = green!5] (9, 0) rectangle (18, 2);
    	        \draw[draw=none, fill = red!5] (18, 0) rectangle (\size, 2);
    	        
    	    }{}
    	    \draw[dashed, black!50] (0, 0) grid (\size, 2);
    	    \begin{scope}[shift={({-0.5 + sqrt(3) / 6}, {1/2-1/3})}]
    	        \pgfmathtruncatemacro{\listlength}{5}
        	    \foreach \x[count=\i] in {8, 7, 4, 3, 1}
        	    {
        	        \pgfmathtruncatemacro{\rowoffset}{mod(\i - 1, 2)}
        	        \pgfmathsetmacro{\xposition}{max(\listlength - \i, \x - ((\i >= \listlength) + (\tick - 1)) * (\tick > 0)}
                    \ifthenelse{\i=1}
                    {
            	        \begin{scope}[shift={({1 - sqrt(3) / 6 + sqrt(3) / 24}, {1/2-1/6})}]
            	            \draw[draw = none, fill = blue!20] (\xposition, 0) circle (0.5cm);
            	        \end{scope}
                    }{}
        	        \begin{scope}[shift = {(\xposition + 1, {\rowoffset * (\tick > 0)})}, scale = \toscale]
                        \import{images/}{blocks.3.blue}
                    \end{scope}
                }
                \foreach \x[count=\i] in {11, 14, 16, 17}
                {
                    \pgfmathtruncatemacro{\delay}{(\i == 2) * (\tick > 0) + (\i == 3) * (\tick > 1) + (\i == 4) * ((\tick > 0) + (\tick > 2))}
        	        \begin{scope}[shift = {(\x - \tick + \delay + 1, 0)}]
        	            \begin{scope}[scale = \toscale]
                            \import{images/}{blocks.3.blue}
                        \end{scope}
                    \end{scope}
                }
            \end{scope}
    	    \begin{scope}[shift={({0.5 - sqrt(3) / 6}, {1/2-1/3})}]
    	        \pgfmathtruncatemacro{\listlength}{4}
        	    \foreach \x[count=\i] in {18, 20, 21, 22}
        	    {
        	        \pgfmathtruncatemacro{\rowoffset}{mod(\i - 1, 2)}
        	        \pgfmathsetmacro{\xposition}{min(\size - \listlength + \i - 1, \x + ((\i < 2) + (\tick - 1)) * (\tick > 0))}
                    \ifthenelse{\i=1}
                    {
            	        \begin{scope}[shift={({0 + sqrt(3) / 6 - sqrt(3) / 24}, {1/2-1/6})}]
            	            \draw[draw = none, fill = red!20] (\xposition, 0) circle (0.5cm);
            	        \end{scope}
                    }{}
        	        \begin{scope}[shift = {(\xposition, {\rowoffset * (\tick > 0)})}, scale=\toscale]
                        \import{images/}{blocks.3.red}
                    \end{scope}
                }
                \foreach \x[count=\i] in {9, 12, 13}
                {
                    \pgfmathtruncatemacro{\delay}{(\i == 2) * (\tick > 1) + (\tick > 0)}
        	        \begin{scope}[shift = {(\x + \tick - \delay, {(\tick > 0)})}, scale = \toscale]
                        \import{images/}{blocks.3.red}
                    \end{scope}
                }
            \end{scope}
        \end{scope}
	\end{tikzpicture}}
		\caption{$t=\tick$}
		\label{sub:Example.Complex.\tick}
	\end{subfigure}
	}
    \caption{We illustrate $5$ time steps of a run of \cref{algo:general.movement} on a non-normal initial configuration. The maximal normal sub-configuration $\mathcal{S}$ is highlighted in green \protect\tikz{\protect\draw[dashed, ultra thin, fill = green!20] (0,0) rectangle (0.28, 0.28);}. The left-hand and right-hand sides of $\mathcal{S}^c$ are highlighted in blue \protect\tikz{\protect\draw[dashed, ultra thin, fill = blue!20] (0,0) rectangle (0.28, 0.28);}  and red \protect\tikz{\protect\draw[dashed, ultra thin, fill = red!20] (0,0) rectangle (0.28, 0.28);}  respectively. The blue $\mathcal{P}^B_{b_0}$ and red $\mathcal{P}^R_{r_0}$ \textbf{anchor agents} are circled in blue and red respectively.}
    \label{figure:subnormalalgoexample}
    \end{figure}
    
    \FloatBarrier
    We will show that \cref{algo:general.movement} is an optimal sorting algorithm.
    
    \begin{lemma}\label{lemma:Alternating-Shift-Algo-runtime}
        If $\mathcal{S} = \varnothing$, the makespan of  \cref{algo:general.movement} is exactly
        \begin{equation*}
            f_{max}^* + \mathcal{V}^*
        \end{equation*}
    \end{lemma}
    
    \begin{proof}
        \cref{lemma:boundextended} shows $f_{max}^* + \mathcal{V}^*$ is a lower bound on the makespan, so we just need to prove that it is also an upper bound.
    
        If $\mathcal{S} = \varnothing$, then at time $t=0$ all the agents split between the two rows in an alternating fashion (see \cref{figure:subnormalalgoexample}). In every subsequent time step any agent that has not reached its final sorted position moves horizontally along its row. Thus, each agent $A \in \mathcal{S}^c$ takes at most $front(A)+1 \leq f^*(A)+1$ time steps to reach its final position in the sorted configuration.
        
        We split the proof into cases.
        
        Case 1: There are only red $f^*$-critical agents.  Recalling \cref{definition:anchoragents}, we infer that the red anchor agent is necessarily $f^*$-critical. Denote this agent $W$. 
        
        Suppose $W$ moves horizontally (i.e., rightwards) at time $t=0$. It will continue to do so at every subsequent time-step, thus reaching its desired column in $front(W) = f^*(W)$ time steps. Note that if $W$ moves horizontally at time $t=0$ according to \cref{algo:general.movement}, it is necessarily the \textbf{only} red $f^*$-critical agent, because it has an empty space in front of it that no other agent counts towards its $f^*$ value. Thus in this case, $\mathcal{V}^*=0$.
        
        Suppose on the other hand that $W$ doesn't move horizontally at time $t=0$. Then, since $W$ is a critical agent, $\mathcal{V}^*=1$, and $W$ will reach its desired column in $front(W) + 1 = f^*(W)+\mathcal{V}^*$ time steps.
        
        Every other agent $A$ reaches its desired column in at most $f^*(A) + 1 \leq f^*(W) + \mathcal{V}^*$ time steps. Thus \cref{algo:general.movement}'s makespan is at most $f_{max}^* + \mathcal{V}^*$.
        
        Case 2: There are only blue  $f^*$-critical agents. This is the same as case 1.
        
        Case 3: There are both red and blue $f^*$-critical agents. Since when $\mathcal{S} = \varnothing$, red and blue agents both move independently on disjoint sets of columns, thus we may combine the arguments of Case 1 and Case 2 to infer that the makespan is $f_{max}^* + \mathcal{V}^*$.
    \end{proof}
    
    \begin{observation}\label{observation:fcriticalalwaysinS}
    If $\mathcal{S} \neq \varnothing$ then any $f^*$-critical agent is necessarily in a column of $S$ at time $t=0$.
    \end{observation}
    
    \begin{proof}
    Suppose there is an $f^*$-critical agent $W$ in a column of $\mathcal{S}^c$. Assume without loss of generality that $W$ is red. Since $\mathcal{S} \neq \varnothing$, there is necessarily a red agent $W'$ in a column of $\mathcal{S}$ with smaller $x$-coordinate than $W$. We have that $|front(W')| > |front(W)|$, since everything that is in front of $W$ is also in front of $W'$, but $front(W')$ contains a blue-occupied column between $W$ and $W'$ which is not in $front(W')$. By definition $f^*(W') \geq |front(W')| > |front(W)| = f^*(W)$, a contradiction to the assumption that $W$ is a critical agent.
    \end{proof}

    \begin{proposition}
    \label{proposition:optimalgeneralmovement}
    In every (normal or non-normal) configuration, \cref{algo:general.movement} completes in exactly
        \begin{equation*}
            f_{max}^* + \mathcal{V}^*
        \end{equation*}
    time steps.
    \end{proposition}
    
    \begin{proof}
    If $\mathcal{S} = \varnothing$, the claim follows from \cref{lemma:Alternating-Shift-Algo-runtime}. We assume therefore that $\mathcal{S} \neq \varnothing$.
    
    Note that at every time step starting at $t = 1$ all agents execute the same logic: red agents move right and blue agents move left whenever they are unobstructed.
    
    In \cref{algo:general.movement}, after time $t=0$, any blue or red agent initialized in the columns of $\mathcal{S}^c$ can move in its desired direction until it settles in its final column in the physical sorting. Furthermore, agents initialized in the columns of $\mathcal{S}$ need at least three time ticks to reach any column of $\mathcal{S}^c$ (see \cref{figure:subnormalalgoexample}). Therefore, agents initialized in  $\mathcal{S}$'s columns can never catch up with agents initialized in $\mathcal{S}^c$ before the  $\mathcal{S}^c$ agents reach their final sorted position. In other words, agents in $\mathcal{S}^c$ never obstruct agents in  $\mathcal{S}$. Moreover, the agents initialized in $\mathcal{S}$ arrive at their final sorted position strictly later than agents in $\mathcal{S}^c$. 
    
    Let $r$ be the number of red agents in the columns of $\mathcal{S}^c$. Note that no agent initialized in  $\mathcal{S}$ will ever enter the rightmost $r$ columns, since these will be occupied by red $\mathcal{S}^c$ agents and \cref{algo:general.movement} does not let two red agents occupy the same column at any time step. Analogously, let $b$ be the number of blue agents in the columns of $\mathcal{S}^c$. Then no agent initialized in  $\mathcal{S}$ will ever enter the leftmost $b$ columns. 
    
    Let $\mathcal{C}^*$ be our initial agent configuration, and let $\mathcal{C}$ be the initial configuration where we delete all agents in the $\mathcal{S}^c$ columns and delete the rightmost $r$ columns and the leftmost $b$ columns. It is not difficult to show from the previous two paragraphs that the makespan of \cref{algo:general.movement} on $\mathcal{C}^*$ is the same as the makespan of \cref{algo:trivial.centralized} on $\mathcal{C}$. 
    
    According to  \cref{observation:fcriticalalwaysinS}, all critical agents in the configuration $\mathcal{C}^*$ are initialized in $\mathcal{S}$. Let $A$ be any red agent initialized in $\mathcal{S}$. Every red agent that we remove from $\mathcal{C}^*$ to create $\mathcal{C}$ (i.e., red agents in $\mathcal{C}^* \setminus \mathcal{C}$) increases $|front(A)|$ by 1. Every column we remove from the right-hand side decreases $|front(A)|$ by $1$ and offsets this. Thus the value of $|front(A)|$ does not change between $\mathcal{C}$ and $\mathcal{C}^*$. The size of $front$ similarly remains unchanged for every blue agent initialized in $\mathcal{S}$. $|back(A)|$ remains unchanged for red and blue agents by definition.  
    
    Consequently, $f_{max}^* + \mathcal{V}^*$ over $\mathcal{C}^*$ is equal to $f_{max} + \mathcal{V}$ over $\mathcal{C}$. Thus the makespan of \cref{algo:general.movement} over $\mathcal{C}^*$ is $f_{max}^* + \mathcal{V}^*$.
    \end{proof}
    
    In conclusion, the makespan of any sorting algorithm over any initial configuration is at least $f_{max}^* + \mathcal{V}^*$. This lower bound is precise; \cref{algo:general.movement} meets it exactly. This generalizes   \cref{theorem:makespanresult} to non-normal configurations.
    
    \begin{theorem}
    The makespan of any sorting algorithm over a given (normal or non-normal) initial configuration is at least $f_{max}^* + \mathcal{V}^*$. This lower bound is precise;  \cref{algo:general.movement} meets it exactly. 
    \end{theorem}

\section{Discussion}

We studied the problem of sorting vehicles or ``agents'' on a horizontal two-row highway, sending red vehicles to the right and blue vehicles to the left as quickly as possible. We derived an exact lower bound for the amount of time it takes an arbitrary configuration of such vehicles starting at the bottom row to arrive at a sorted configuration, and presented an optimal sorting algorithm that attains this lower bound.

The instance-optimal algorithm we presented for sorting normal configurations (\cref{algo:trivial.centralized}) requires global knowledge of the initial agent configuration to compute $f_{max}$, as the value of $f_{max}$ determines which color of agent it raises to the upper row. Consider the algorithm that, instead of computing $f_{max}$, simply raises all the \textit{blue} agents to the upper row, and otherwise proceeds the same as \cref{algo:trivial.centralized}. This is a straightforward algorithm that requires no global knowledge; in fact, it can be implemented by decentralized agents with local sensing and no communication (see Appendix A, \cref{algo:trivial.distributed}). \cref{theorem:makespanresult} shows that this algorithm is at most $1$ time step slower than \cref{algo:trivial.centralized}. \cref{fig:comparison run} in Appendix A shows an example run of both algorithms. The authors find it significant that the instance-optimal strategy for this problem can be approximated by a very simple, decentralized and local sensing-based algorithm. 

It is not clear whether a simple decentralized strategy exists for non-normal configurations. In our general sorting algorithm (\cref{algo:general.movement}),  agents must know whether they are in the maximal subnormal configuration or outside of it to determine their movements, and so a local decentralized algorithm with nearly identical performance is more difficult to conceive of. 

In future work, it will be interesting to consider the following extensions of our problem:

\begin{enumerate}
    \item Settings where there are more than two rows which the agents may use, and where agents are initialized on arbitrary rows and columns.
    \item General permutations. Suppose that instead of just $2$ colors, there are $k$ different colors. What is the optimal way to sort the agents, such that all agents of color $i$ are between those of color $i-1$ and $i+1$? 
    \item Agents moving at different velocities, e.g., a grid square every $2$ or $4$ time steps instead of $1$.
\end{enumerate}

Physical sorting problems such as the one described in this paper force algorithm designers to take into account a number of factors that are not present in more traditional settings, such as physical motion and collision avoidance. When compared to traditional combinatorial algorithms, we believe that there is a lot of room for further theoretical developments in this domain, as even results that at a glance might seem straightforward currently require specialized analysis.


\section*{Acknowledgments}
The research presented in this paper was partially funded by the The Israeli Smart Transportation Research Center (ISTRC).


\clearpage

\begin{appendix}
    \section{An almost optimal distributed solution}
    The proposed centralized solution can be easily parallelized to produce an almost trivial decentralized algorithm, that, nonetheless, sorts agents in only a single tick slower than the instance-optimal \cref{algo:trivial.centralized}. Every agent $A$ has a memory bit $d$ which is used to track whether the current time step is $0$ or not. At time $0$, all the blue agents move to the upper row. In \cref{algo:trivial.distributed}, we describe what each (decentralized) mobile agent does at every time tick $t$:
    
	\begin{algorithm}
		\SetAlgoLined
		\KwResult{A sorted agent configuration.}
		\kwInit{memory bit $d \gets 0$}
		
		\uIf{I am a blue agent}
		{
			\uIf{$d = 0$}{
				move to the second row\;
			}
			\Else{
				\If{adjacent location to the left is empty} {
					move one step left\;
				}
			}
		}
		\Else
		{
			\If{adjacent location to the right is empty} {
				move one step right\;
			}
		}
		
		$d \gets 1$\;
		\caption{Almost optimal decentralized sorting algorithm for an unsorted initial normal configuration. }
		\label{algo:trivial.distributed}
	\end{algorithm}		
	
\begin{figure}
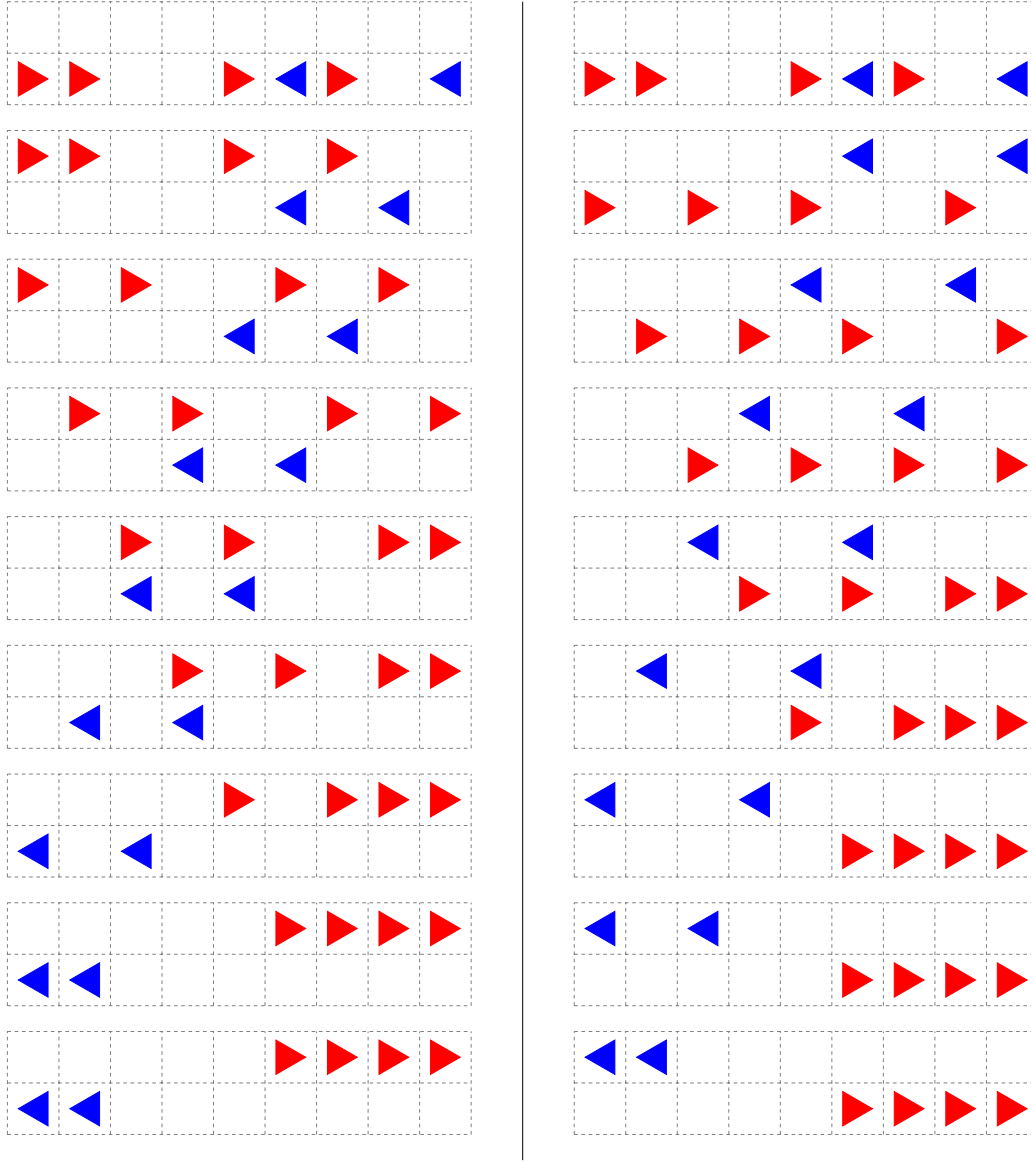

        \centering
        \resizebox{\textwidth}{!}{	\begin{tikzpicture}
	    \pgfmathsetmacro{\steplength}{1.5}
	    \pgfmathsetmacro{\size}{9}
	    \pgfmathsetmacro{\height}{2}
	    \pgfmathsetmacro{\ygridoffset}{-\height - 0.5}
	    \pgfmathsetmacro{\split}{2}
	    \begin{scope}[scale=\steplength]
	        \foreach \tick in {0, ..., 8}
	        {
	            \pgfmathsetmacro{\yoffset}{\ygridoffset * \tick}
	            \begin{scope}[yshift = \yoffset cm]
            	    \draw[dashed, black!50] (0, 0) grid +(\size, \height);
            	    \draw[dashed, black!50] (\size+\split, 0) grid +(\size, \height);
	            \end{scope}
	        }
            \draw[thick] (\size + \split / 2, \height) -- +(0, \ygridoffset * 9);

    	    \begin{scope}[shift={({-0.5 + sqrt(3) / 6}, {1/2-1/3})}]
        	    \foreach \tick/\x in 
        	        {
        	            0/6, 0/9,
        	            1/6, 1/8,
        	            2/5, 2/7, 
        	            3/4, 3/6,
        	            4/3, 4/5,
        	            5/2, 5/4,
        	            6/1, 6/3,
        	            7/1, 7/2,
        	            8/1, 8/2
        	        }
        	    {
        	        \begin{scope}[shift = {(\x, \ygridoffset * \tick)}, scale = 2/3]
                        \import{images/}{blocks.3.blue}
                    \end{scope}
                }
            \end{scope}
    	    \begin{scope}[shift={({0.5 - sqrt(3) / 6}, {1/2-1/3})}]
        	    \foreach \x in {0, 1, 4, 6}
        	    {
        	        \begin{scope}[xshift = \x cm, scale=2/3]
                        \import{images/}{blocks.3.red}
                    \end{scope}
                }
        	    \foreach \tick/\x in 
        	        {
        	            1/0, 1/1, 1/4, 1/6,
        	            2/0, 2/2, 2/5, 2/7,
        	            3/1, 3/3, 3/6, 3/8,
        	            4/2, 4/4, 4/7, 4/8,
        	            5/3, 5/5, 5/7, 5/8,
        	            6/4, 6/6, 6/7, 6/8,
        	            7/5, 7/6, 7/7, 7/8,
        	            8/5, 8/6, 8/7, 8/8
        	        }
        	    {
        	        \begin{scope}[shift = {(\x, 1 + \ygridoffset * \tick)}, scale=2/3]
                        \import{images/}{blocks.3.red}
                    \end{scope}
                }
            \end{scope}
            
    	    \begin{scope}[shift={({-0.5 + sqrt(3) / 6 + \size + \split}, {1/2-1/3})}]
    	        \foreach \x in {6, 9}
    	        {
        	        \begin{scope}[shift = {(\x, 0)}, scale = 2/3]
                        \import{images/}{blocks.3.blue}
                    \end{scope}
    	        }
        	    \foreach \tick/\x in 
        	        {
        	            1/6, 1/9,%
        	            2/5, 2/8,%
        	            3/4, 3/7,%
        	            4/3, 4/6,%
        	            5/2, 5/5,%
        	            6/1, 6/4,%
        	            7/1, 7/3,%
        	            8/1, 8/2%
        	        }
        	    {
        	        \begin{scope}[shift = {(\x, \ygridoffset * \tick + 1)}, scale = 2/3]
                        \import{images/}{blocks.3.blue}
                    \end{scope}
                }
            \end{scope}
    	    \begin{scope}[shift={({0.5 - sqrt(3) / 6 + \size + \split}, {1/2-1/3})}]
        	    \foreach \tick/\x in 
        	        {
        	            0/0, 0/1, 0/4, 0/6,
        	            1/0, 1/2, 1/4, 1/7,
        	            2/1, 2/3, 2/5, 2/8,
        	            3/2, 3/4, 3/6, 3/8,
        	            4/3, 4/5, 4/7, 4/8,
        	            5/4, 5/6, 5/7, 5/8,
        	            6/5, 6/6, 6/7, 6/8,
        	            7/5, 7/6, 7/7, 7/8,
        	            8/5, 8/6, 8/7, 8/8
        	        }
        	    {
        	        \begin{scope}[shift = {(\x, \ygridoffset * \tick)}, scale=2/3]
                        \import{images/}{blocks.3.red}
                    \end{scope}
                }
            \end{scope}
        \end{scope}
	\end{tikzpicture}}
        \caption{A side-by-side algorithm execution on a given initial grid configuration. (Left) centralized \cref{algo:trivial.centralized}, (right) distributed \cref{algo:trivial.distributed}. The centralized algorithm finishes $1$ time step before the decentralized algorithm ($t=7$ and $t=8$ respectively).}
        \label{fig:comparison run}
    \end{figure}

    Based on the analysis of \cref{theorem:makespanresult}, this algorithm is at most $1$ time step slower than the optimal time for any given initial configuration. The additional time tick is incurred because in an instance-optimal solution, sometimes we must raise the red agents to the second row rather than the blue agents. In \cref{fig:comparison run}, we provide an example run comparing \cref{algo:trivial.centralized} and \cref{algo:trivial.distributed}.
    
    \FloatBarrier
    
\end{appendix}

\bibliographystyle{elsarticle-num} 
\bibliography{biblio}

\end{document}